\documentclass[11pt]{article}

\usepackage{fullpage}
\usepackage{amsmath, amssymb }
\usepackage{graphicx,longtable,wrapfig}

\newtheorem{Th}{Theorem}

\newtheorem{Prop}{Proposition}

\renewcommand{\geq}{\geqslant}

\renewcommand{\le}{\leqslant}
\renewcommand{\ge}{\geqslant}

\newcommand{\Prob}{{\rm P}}
\newcommand{\E}{{\rm E}}
\newcommand{\Host}{{\mathrm{Host}}}

\title{\bfseries Empirical Validation of the Buckley--Osthus Model for the Web Host Graph: Degree and Edge Distributions%
\footnote{Short version of this paper will be published in the proceedings of CIKM~2012 (see \texttt{http://www.cikm2012.org/}).}
}

\author{Maxim~Zhukovskiy\qquad Dmitry~Vinogradov\qquad Yuri~Pritykin\qquad
Liudmila~Ostroumova \\ \qquad Evgeniy~Grechnikov\qquad  Gleb~Gusev\qquad
Pavel~Serdyukov\qquad Andrei~Raigorodskii \\[2mm]
       Yandex\\
       16 Leo Tolstoy St., Moscow, 119021 Russia\\[2mm]
       \texttt{\{zhukmax, dimavin, pritykin, ostroumova-la, grechnik,}\\
       \texttt{gleb57, pavser, raigorodsky\}@yandex-team.ru}
}

\begin{document}

\maketitle

\begin{abstract}

There has been a lot of research on random graph models for large
real-world networks such as those formed by hyperlinks between web
pages in the world wide web. Though largely successful
qualitatively in capturing their key properties, such models may
lack important quantitative characteristics of Internet graphs. While preferential attachment random graph models were shown to be capable of reflecting the degree distribution of the webgraph, their ability to reflect certain aspects of the edge distribution
was not yet well studied.

In this paper, we consider the Buckley--Osthus implementation of preferential attachment and its ability to model the web host graph in two aspects. One is the degree distribution that we observe to follow the power law, as often being the case for real-world graphs. Another one is the two-dimensional edge distribution, the number of edges between vertices of given degrees. We fit a single ``initial attractiveness'' parameter $a$ of the model, first with respect to the degree distribution of the web host graph, and then, absolutely independently, with respect to the edge distribution. Surprisingly, the values of $a$ we obtain turn out to be nearly the same. Therefore the same model with the same value of the parameter $a$ fits very well the two independent and basic aspects of the web host graph. In addition, we demonstrate that other models completely lack the asymptotic behavior of the edge distribution of the web host graph, even when accurately capturing the degree distribution.

To the best of our knowledge, this is the first attempt for a real graph of Internet to describe the distribution of edges between vertices with respect to their degrees.
\end{abstract}


\newcommand{\MExpect}{\mathsf{M}}

\newcommand{\BR}{{Bollob\'as--Riordan}}
\newcommand{\BO}{Buckley--Osthus}
\newcommand{\BA}{Barab\'asi--Albert}
\newcommand{\B}{Buckley--Osthus}
\newcommand{\ER}{Erd\H{o}s--R\'enyi}

\section{Introduction}

The study of the web as a hyperlink graph yields a valuable insight into web algorithms for crawling, searching, and community discovery~\cite{Bonato2005, Efe2000, Kumar2000pods}. Valid random graph models of the web provide methods of generating WWW-like graphs that are significantly smaller and simpler than real WWW graphs, but yet preserve certain key properties of the hyperlink structure of the web. Such artificial graphs could serve as a convenient experimental platform, where new approaches to search, indexing, compression can be evaluated.

Vertices of the \emph{webgraph} correspond to web pages and edges
represent hyperlinks between them. Webgraphs have been extensively studied with respect to many quantitative aspects such as degree distribution, diameter, number of connected components, macroscopic structure, and assortative mixing (e.g., see~\cite{power3, power4, power1,Costa,  power2, Kumar2000pods, DynAndCorrOfTheInternet}).



In this paper, we consider the \emph{web host graph}. Vertices of
this graph are web hosts and edges correspond to hyperlinks between their pages. The web host graph is much smaller than the webgraph, but
is still a very useful resource and an abstraction of the web.
For a lot of purposes, modern search engines consider hosts (and
web sites associated with them) rather than web pages as the
smallest possible entities in the web. Particularly, the smaller size
of the web host graph allows for simpler and more efficient link
analysis useful for web search related tasks.

We study the web host graph from two perspectives.

First, we look at the distribution of degrees of the web host graph vertices. It was shown that degrees of the webgraph vertices, much like in many other real world networks~\cite{Easley2010}, obey the power law
\cite{power3,power4,power1,power2}. Albert et al.\ were the first
to find the power law in the degree distribution of the web pages in
the domain *.nd.edu~\cite{power3}. Not surprisingly, we observe that the degree distribution in the web host graph also follows the power law.

Second, we study how edges in the web host graph are distributed between vertices depending on degrees of these vertices. To the best of our knowledge, this is the first study of this property for graphs of Internet. However, in a reduced form, this notion was previously studied under the name of \emph{degree correlation} or \emph{assortativity} (e.g., see~\cite{power4}). A~convenient way of capturing the degree correlation is by examining the properties of $d_{nn} (d)$, the expected average degree of neighbors of a random vertex with degree $d$. In real-word networks, one often observes $d_{nn}(d) \sim d^{\delta}$ with some $\delta$, which is negative for the webgraph (disassortativity) \cite{DynAndCorrOfTheInternet} and usually positive for social networks (assortativity)~\cite{AssortMixing}. We study a more general property of \emph{the distribution of edges} between vertices with respect to their degrees, that is, the total number of edges $X(d_1,d_2)$ between pairs of vertices with degrees $d_1, d_2$. In fact, one can obviously derive both degree distribution and $d_{nn}(d)$ from the edge distribution: $\#(d) = 1/d\sum_{d_1} X(d_1,d),\, d_{nn}(d) = \frac{\sum_{d_1} d_1 X(d,d_1)}{\sum_{d_1} X(d,d_1)}$. On the other hand, the degree distribution of a real-world graph does not determine its edge distribution, and therefore the latter may be considered as an additional more general aspect of the graph. In fact, there are assortative and disassortative real-world graphs with close power-law degree distributions, and therefore their edge distributions do differ as well.

There are a number of important random graph models whose features
(such as degree distribution, diameter, etc.) are supposed to be
close to those of the real Internet graphs and social networks. Barab\'asi and Albert proposed the most well-known approach \cite{BA} that was realized in various preferential attachment models. Several of them have precise mathematical definitions~--- the Bollob\'as and Riordan model~\cite{BR1}, the Buckley and Osthus model~\cite{BO,Dorog,Drin}, the 
Copying Models~\cite{EGC,Stochastic}, Directed Scale-Free
Graphs~\cite{Scalefree} and the general model of
Cooper and Frieze~\cite{Cooper} (see also~\cite{Stochastic}).
An extensive review of these models can be found elsewhere (e.g., see~\cite{power4,Res}).

We focus on random graph models that allow for mathematical analysis of their properties and thus have to be relatively simple. Bollob\'as and Riordan were the first to propose a precisely defined preferential attachment model, and proved the power law for degree distribution in this model with mathematical rigor. It was shown that the number of vertices with degree $d$ decreases as $d^{-\gamma}$ with $\gamma = 3$~\cite{BR2}. On the other hand, Barab\'asi and Albert empirically estimated this exponent in the real webgraph to be approximately $2.1 \pm 0.1$\quad\cite{BA}. As we show in this paper, the parameter $\gamma$ for the web host graph is also far from~3. Therefore the model of Bollob\'as and Riordan is not realistic for both the webgraph and the web host graph. The random graph model of Buckley and Osthus, which is a generalization of the Bollob\'as--Riordan model, solves this problem. Namely, the Buckley--Osthus model depends on a parameter $a$ called \emph{initial attractiveness}. For an integer value of $a$, Buckley and Osthus proved theoretically that the degree distribution in this model also follows the power law with the exponent $-2 - a$\quad\cite{BO}.

Recently Grechnikov obtained two results concerning the \BO\ model~\cite{Gr2}. First, he extended the result from~\cite{BO} about the degree distribution to an arbitrary positive $a$, not necessarily integer. Second,
he obtained an accurate asymptotic estimate for the edge distribution
for growing
degrees as an explicit formula depending on~$a$ as a parameter.

We expect the \BO\ model to be a good approximation of the web host graph to a certain extent. Relying on the both aforementioned theoretical results concerning two different aspects of the \BO\ model, we find the best fit of the initial attractiveness parameter~$a$ for the real web host graph assuming it is generated in the model. First, we choose the value of the parameter $a$ so that the exponent in the power law for the real web host graph is close to $-2-a$. In a second approach, completely independent from the first one, we estimate $a$ using the best fit
of the formula from~\cite{Gr2} for the edge distribution in the \BO\ model to the really observed edge distribution in the web host graph.
Surprisingly, in both cases we find out that the model agrees very well with the real graph with the same value of $a\approx 0.3$. In other words, this very same model with $a \approx 0.3$ accurately approximates two completely different and a priori independent basic aspects of the web host graph, degree and edge distributions (and therefore also assortativity). This is especially impressive as the model itself is very simple and has only a single degree of freedom, namely the initial attractiveness parameter~$a$. Note that we were able to describe the distribution of edges in a real web host graph only using the model for this graph and properties of this model obtained theoretically. Without the model, to come up with the asymptotics of the edge distribution would be really hard, if not impossible.


We compare the \BO\ model and its ability to describe the real web host graph with other random graph models. We focus on the models that generate graphs with the degree distribution following the power law. We demonstrate that degrees do not immediately relate to edges between vertices given their degrees. Even after being fit to the web host graph with respect to degree distribution, the models are not able to capture the edge distribution and in fact completely lack the asymptotic behavior of this edge distribution observed in the web host graph and in the \BO\ model.



To sum up, contributions of this paper are the following:

\begin{itemize}

\item We have studied the distribution of degrees and the distribution of edges between vertices given their degrees for the web host graph.

\item Relying on previous theoretical study, we demonstrated that the web host graph corresponds very well to the \BO\ random graph model with the initial attractiveness parameter $a\approx 0.3$ with respect to the degree distribution, the edge distribution and (consequently) the assortativity. We obtained the same value of the parameter two times by fitting the model with respect to the both independent quantitative aspects: the degree and the edge distributions.

\item We generated graphs in the \BO\ model and empirically examined their theoretically proved properties in practice. We also showed that some other random graph models fail to capture the edge distribution of the web host graph though may successfully capture the degree distribution.
\end{itemize}

To the best of our knowledge, this is the first attempt to empirically validate the Buckley--Osthus model on the real web data. Moreover, this is the first attempt to rigorously study the distribution of edges between vertices with respect to their degrees for graphs of Internet.

The remainder of the paper is organized as follows.
Section~\ref{Models} is a short review on random graph models. In particular, in Section~\ref{Theory} we describe the theoretical properties of the \BO\ model critical for our experiments with the web host graph. We describe the experimental part of the work in Sections~\ref{Results} and~\ref{Simulation}. In Section~\ref{Results}, we report the results of the approximation of the web host graph by the \BO\ model with respect to degree and edge distributions. In Section~\ref{Simulation}, we compare it with approximations by other models. In Section~\ref{Conclusion} we discuss potential applications and future work.

\section{Random Webgraph Models}
\label{Models}

One possible theoretical approach to what the model of a webgraph might be is the mathematical concept of random graphs. The essence of this approach is in the idea of a webgraph developing stochastically. Once the rules or the parameters of this stochastic process are precisely specified, a random graph may obtain (sometimes unexpected) stable properties, in spite of the stochastic nature of its formation. Some of such properties may reflect those of the real webgraph rather accurately.

There have been a lot of attempts to model the hyperlink graph of the web as a random graph. Probably the simplest (even regardless of the webgraph) are random graphs of the \ER\ model, where a graph is constructed by creating a fixed number of vertices and a fixed number of edges drawn independently uniformly at random over pairs of vertices. However, this model is not suitable for the webgraph (as well as the web host graph) as it lacks scalability, that is, does not have a power law degree distribution.

In 1999, Barab\'asi and Albert~\cite{BA} observed that the degree distribution of the real webgraph follows the power law with the exponent approximately equal to $-2.1$. They proposed a concept of preferential attachment that explained the phenomenon. The basic underlying idea is the following. A graph is constructed with a random process. At each step of the process, a new vertex is added and a fixed number of edges are added from the new vertex to randomly chosen already existing vertices. Vertices with higher degree acquire new edges with higher probability that linearly depends on their degree (``rich get richer'').

\subsection{Preferential attachment models}

The general idea of preferential attachment obtained a precise mathematical formulation in the model of Bollob\'as and Riordan~\cite{BR2} defined in the following way. We construct a series of graphs (Markov chain) $G_m^n, n=1,2,\ldots$, with $n$ vertices and $mn$ edges, where $m\in\mathbb{Z}$ is a fixed number. Let us consider the case $m = 1$ first.
Let $G_1^1$ be a graph consisting of one vertex with a
self-loop. A graph $G_1^{t}$ is obtained from $G_1^{t-1}$ by
adding a vertex $t$ and an edge from $t$ to a vertex $i$, where
$i$ is chosen randomly within the existing vertices with the
following probability distribution:
$$
\Prob(i=s) = \
\begin{cases}
d_{G_1^{t-1}}(s)/(2t-1) & \text{if } 1 \le s \le t-1,\cr \noalign{\smallskip}
{1/(2t-1)} &\text{if }  s=t, \cr
\end{cases}
$$
where $d_{G_1^t}(s)$ denotes the degree of the vertex~$s$ in 
$G_1^t$.
A graph $G_m^n$ is constructed from $G_1^{mn}$ by merging the vertices $1, \ldots, m$ into the vertex $1$ of the new graph, merging the vertices $m+1, \ldots, 2m$ into the vertex 2 of the new graph, etc. Note that one can consider the variant of the model with directed graphs: in this case, an edge between the vertices $i$ and $j$ goes from $i$ to $j$ if $i>j$.

The \BR\ model accurately captures some of the key properties of different real-world graphs. For instance, ``small-world phenomenon'' of many real-world networks, i.e., a surprisingly small diameter, is also
observed in the model. Bollob\'as and Riordan proved that indeed the diameter of $G_m^n$ is about $\frac{\log n}{\log \log n}$ for large~$n$\quad\cite{BR1}. They also showed that the degree distribution of $G_m^n$ obeys the power law: the number of vertices with degree $d$ in the model is well approximated by $d^{-\gamma}$, with $\gamma = 3$\quad\cite{BR2}. However, this disagrees with the webgraph where the estimate $\gamma_{WWW} = 2.1\pm 0.1$ was observed~\cite{BA}. This means that even though the \BR\ model is similar in some aspects to real graphs of Internet qualitatively, it needs to be refined to better capture the reality quantitatively. In this work, we estimate the value of $\gamma_{\Host}$ in the power law for the web host graph to be approximately $2.276 \pm 0.001$ (see Section~\ref{CumDistr}).





A possible approach for such a refinement is the model proposed independently by two groups of researchers~\cite{Dorog, Drin}.
They proposed to extend the model with a parameter called
\emph{initial attractiveness} of a vertex, a positive constant that does not depend on degree. Later Buckley and Osthus gave an explicit construction of this model~\cite{BO}. The degree distribution of a \BO\ graph also obeys the power law, but now varying the value of $a$ in the definition of the model, one can tune the exponent~$\gamma$ in the power law of the resulting graph.

More specifically, the model generates
a series of graphs $H_{a,m}^n, n=1,2,\ldots,$ with $n$ vertices and $mn$ edges, where $m\in \mathbb{Z}$ is a fixed number. The definition of $H_{a,1}^n$ recapitulates the definition of $G_m^n$, and the only difference is that the probability of a newly added edge in $H_{a,1}^n$ equals
$$
\Prob(i=s) = \
\begin{cases}
\frac{d_{H_{a,1}^{t-1}}(s) + a - 1}{(a+1)t-1} & \text{if } 1 \le s \le t-1,\cr \noalign{\smallskip}
\frac{a}{(a+1)t-1} &\text{if }  s=t. \cr
\end{cases}
$$
A graph $H_{a,m}^n$ is obtained from $H_{a,1}^{mn}$ in the same
manner as $G_m^n$ from $G_1^{mn}$. Note that for $a=1$, we obtain the initial Bollob\'as--Riordan model $G_m^n$. For an integer $a$, Buckley and Osthus proved \cite{BO} that the degree distribution of a random graph in the model follows the power law with $\gamma = 2+a$.


Previously, the \BO\ model has not been compared with real graphs.
In Section~\ref{Theory}, we present further properties of the \BO\ model obtained recently and then use them in Section~\ref{Results} for comparison of this model with the real web host graph.

\subsection{Properties of the \BO\ Model}
\label{Theory}

In this section, we present recent theoretical results on degree and edge distributions of the \BO\ random graph model~\cite{Gr2}.
Our experiments with real graphs (Section~\ref{Results}) are based on the results from this section.

\subsubsection{Degree distribution}

The following theorems show the dependence of the degree distribution in the \BO\ model on the initial attractiveness parameter~$a$ and generalize the results of \cite{BO}.

For a given pair of functions $f,g$, we say
that $f(n) = O(g(n))$ if there exists a positive real number $C$ such that $|f(n)| \le C
g(n)$ for sufficiently large~$n$. We also say that $f(n) =
o(g(n))$ if $f$ is dominated by $g$ asymptotically and $f(n) =
\omega(g(n))$ if $f$ dominates $g$ asymptotically.


Let $\#_a(d,n)$ be the number of vertices with degree $d$
in the model $H_{a,m}^n$. We denote by $\E X$ the expectation of a random variable $X$ .

\begin{Th}[\cite{Gr2}]
\label{BO-degree-mean}
For $d \ge m$ and for every fixed positive $a$,
$$
\E \left( \#_a(d,n) \right) =\frac{{\rm B}(d-m+ma,a+2)}{{\rm B}(ma,a+1)}n+O\left(\frac {1}{d}\right).
$$
\end{Th}
Here ${\rm B}(x,y)$ is the beta function. Note that
$$
\frac{{\rm B}(d-m+ma, a+2)}{{\rm B}(ma,a+1)}\sim C d^{-2-a}
$$
as $d \to \infty$ with $C = (a+1)\frac{\Gamma(ma+a+1)}{\Gamma(ma)}$,
where $\Gamma(x)$ is the gamma function (an extension of the factorial function).

We say that a certain property holds \textbf{whp} (with high probability) if the probability of this property tends to $1$ as $n\rightarrow\infty$. The following concentration result shows that the degree distribution obeys the power law with $\gamma = 2+a$.
\begin{Th}[\cite{Gr2}]
\label{BO-degree-var}
Consider $d \ge m$ to be the value of a function of $n$ and $\psi(n)$ to be a function tending to infinity arbitrarily slowly. Then \textbf{whp}
we have
$$
\left| \#_a(d,n)-\frac{{\rm B}(d-m+ma,a+2)}{{\rm
B}(ma,a+1)}n \right|\le
$$
$$
\le \left(\sqrt{d^{-a-2}n}+d^{-1}\right)\psi(n).
$$
\end{Th}

In contrast with the result of \cite{BO}, $a$ is not necessarily integer here. Roughly speaking, Theorems~\ref{BO-degree-mean} and~\ref{BO-degree-var} imply that for large $d$, we have
\begin{equation}
\label{Asymptotics_1}
\#_a(d,n) \sim b_1 d^{-2-a}n
\end{equation}
with some constant~$b_1$ in an appropriate range of degrees $d$.




\subsubsection{Edges between vertices of given degrees}
\label{EdgesTheory}

In this subsection, we report the results capturing the behaviour of $X_a(d_1, d_2, n)$, the total number of edges between vertices of degree $d_1$ and vertices of degree $d_2$ in a \BO\ graph.
For $d_1 = d_2$, we count every edge twice, but we do not count self-loops. We use this function in Section~\ref{Results} comparing the web host graph with some random graph models including the \BO\ model.

The number of edges between vertices of given degrees in the \BO\ model can be estimated in the following way.

\begin{Th}[\cite{Gr2}]
\label{BO-edges-mean}
Consider $d_1, d_2$ to be the values of two functions of $n$ tending to infinity as $n$ grows. Then
$$
\E X_a(d_1,d_2,n)=c_a(d_1,d_2)n+O(1),
$$
where
$$
c_a(d_1,d_2) = ma(a+1)\frac{\Gamma(ma+a+1)}{\Gamma(ma)}\frac{(d_1+d_2)^{1-a}}{d_1^2 d_2^2} \times
$$
$$
\times \left(1+O\left(\frac1{d_1}+\frac1{d_2}+\frac{d_1d_2}{(d_1+d_2)^2}\right)\right).
$$
\end{Th}

The following theorem is a concentration result.

\begin{Th}[\cite{Gr2}]
\label{BO-edges-var}
Let $c>0$. Then
$$
\Prob \left( |X_a(d_1,d_2,n)-\E X_a(d_1,d_2,n)|\ge c(d_1+d_2)\sqrt{mn} \right) \le
$$
$$
\le 2\exp\left(-\frac{c^2}8\right)\,.
$$
In particular, for an arbitrary function $c(n)$ tending to infinity as $n$ grows we have \textbf{whp} $|X-\E X|<c(n)(d_1+d_2)\sqrt{mn}$.
\end{Th}

Thus it follows that $\E X_a(d_1,d_2,n)$ behaves as
\begin{equation}\label{Asymptotics_2}
(d_1+d_2)^{1-a} d_1^{-2}d_2^{-2} n
\end{equation}
if the ratio $\max(d_1,d_2) / \min(d_1,d_2)$ is sufficiently large
(otherwise 
this formula does not capture the asymptotic behavior).


In Section~\ref{Results}, we also use the fact that the number of loops and multiple edges in the \BO\ random graph is considerably smaller than the total number of edges. To be more precise, the following statement holds.

\begin{Prop}
\label{multiple}
For every $0<a<1$ we have
$$
\E N(\text{\emph{loops in }}H_{a,m}^n) = O\left(\ln n\right),
$$
$$
\E N(\text{\emph{multiple edges in }}H_{a,m}^n) =
O\left(n^{1-a}\right).
$$
\end{Prop}

Proposition~\ref{multiple} is proved in Appendix.
Here we denote by $N(\text{loops in }H_{a,m}^n)$ the number of loops in $H_{a,m}^n$ and denote by $N(\text{multiple edges in }H_{a,m}^n)$ the number of multiple edges in the random graph. Recall that the total number of edges in $H_{a,m}^n$ is $m n$ and dominates both $n^{1-a}$ and $\ln n$.

\subsection{Other results related to edge distribution}

The distribution of edges between vertices with respect to their degrees is closely related to
an interesting quantitative characteristic of graphs called \emph{assortativity}, or \emph{degree correlation} \cite{power4, Costa, AssortMixing, Park, DynAndCorrOfTheInternet}.
Informally, a graph is \emph{assortative} (has a positive degree correlation) if vertices
of high degree tend to connect with vertices of high degree. On the other hand, a graph is
\emph{disassortative} (has a negative degree correlation) if vertices of high degree tend to connect with vertices of low degree.
A convenient way of capturing the degree correlation is by examining the properties of $d_{nn} (d)$, the average degree of neighbors of a vertex with degree $d$ (first average over neighbors of a vertex, and then over vertices of degree $d$). In real-word networks, often $d_{nn}(d) \sim d^{\delta}$ with some $\delta$, which is negative for the webgraph (disassortativity) \cite{DynAndCorrOfTheInternet} and usually positive for social networks (assortativity)~\cite{AssortMixing}. Disassortativity of protein networks was studied in~\cite{Maslov}.

The function $X(d_1,d_2)$ of edge distribution, the number of edges between vertices of degrees $d_1, d_2$, may be considered as a generalization of~$d_{nn}(d)$. In fact, the latter can be restored from the former:
\begin{equation*}
d_{nn}(d) = \frac{\sum_{d_1} d_1 X(d,d_1)}{\sum_{d_1} X(d,d_1)}.
\end{equation*}

As mentioned in \cite{AssortMixing} and \cite{DynAndCorrOfTheInternet}, networks in the \BA\ model \cite{AssortBA} have $d_{nn}(d) \approx \mathrm{const}$ and thus do not demonstrate assortative mixing. However, it can be shown experimentally that a graph in the Buckley--Osthus model, that is a generalization of the \BA\ model, may demonstrate assortativity (for $a>1$) or disassortativity (for $a<1$). In particular, we compare the web host graph and the \BO\ model with $a\approx 0.3$ with respect to the function $d_{nn}(d)$ in Section~\ref{Simulation} and find them to be close to each other (see Fig. \ref{fig:assortativity}).

It was claimed in \cite{MSZ} that the negative degree correlation may be explained by the model where a graph is chosen uniformly at random from the set of all graphs with a prescribed power-law degree distribution without multiple edges. The authors stated that the resulting graph will have a negative degree correlation with high probability, for vertices of high degree are forced to connect each other rarely, or otherwise multiple edges will be more likely to appear.
The authors of \cite{Park} obtain some theoretical results for a similar model. They also argue that the graph of Internet is disassortative. In \cite{CCP} the assortative co-authorship graph is modeled. The proposed model is based on preferential attachment, with an additional idea of adding new links between already existing vertices chosen based on their degrees.
This idea can be utilized for modeling both assortative and disassortative graphs. However, in contrast with the \BO\ model, these models are not based on any natural rules that would explain the underlying laws of graph formation. They are rather specific and thus may be suspected in ``overfitting'' when approximating real graphs.

\section{Experiments on the web host graph}
\label{Results}

\subsection{Preliminaries}
\label{Preliminaries}

Let us consider a random graph in the \BO\ model~$H_{a,m}^n$. For
simplicity, we ignore edge directions, merge multiple
edges and remove loops. Due to Proposition~\ref{multiple}, the
difference between the obtained graph $H$ and the initial one is
not important for us and Theorems \ref{BO-degree-mean}, \ref{BO-degree-var}, \ref{BO-edges-mean}, \ref{BO-edges-var} are still applicable to~$H$.

In what follows,
we denote by $\#(d)$ the number of vertices of degree~$d$ and by
$X(d_1,d_2)$ the total number of edges between all vertices of
degree~$d_1$ and all vertices of degree~$d_2$. The following two
properties follow from equations (\ref{Asymptotics_1}), (\ref{Asymptotics_2}).
\begin{itemize}
\item[1)] The function $\#(d)$ is approximated well by
\begin{equation}
\label{v_estimate}
b_1d^{-2-a}
\end{equation}
for some constant $b_1$ in an appropriate range of degrees.

\item[2)] The number $X(d_1,d_2)$ of
edges between vertices of degrees $d_1,d_2$ is approximated well
by the function

\begin{equation}
\label{X_estimate}
b_2(d_1+d_2)^{1-a}d_1^{-2}d_2^{-2}
\end{equation}

for some constant $b_2$ in an appropriate range of degrees.
\end{itemize}


For the web host graph (undirected, without loops or multiple edges, see Section~\ref{Data} for details) we define $\#_{\Host}(d)$ and $X_{\Host}(d_1,d_2)$ in exactly the same manner as for a random graph~$H$. Each of these two functions can be considered as an empiric density function of some distribution. Indeed, let $\xi$ be the degree of a random vertex and $\psi$ be the ordered pair of degrees of vertices adjacent to a random edge (here the order of the vertices is also chosen randomly). Then the function $\#_{\Host}$ is the empirical density function of the random quantity $\xi$ and $X_{\Host}$ is that of the random vector~$\psi$.

It is known
that as $d$ grows, the variation of the function $\#_{\Host}(d)$
may dominate its mean~\cite{PowerLaw}, see figures in \cite{power1}. The same might by true for the function $X_{\Host}(d_1,d_2)$
as $d_1$ and $d_2$ grow. Therefore it is more convenient, in
particular less vulnerable to fluctuations of the data, to study
the corresponding distribution functions instead of the density
functions.

To that end, we consider the following \emph{cumulative} functions:
\begin{equation}
\begin{split}
&\widetilde{\#}_{\Host}(d) = \sum_{j>d}{\#_{\Host}(j)},\\ &\widetilde{X}_{\Host}(d_1,d_2)= \sum_{j_1\geq j_2,j_1>d_{\max}, j_2>d_{\min}} X_{\Host}(j_1,j_2),\\
&\widetilde{\rho}_{\Host}(d_1,d_2)=
\frac{\widetilde{X}_{\Host}(d_1,d_2)}{\widetilde{\#}_{\Host}(d_1) \widetilde{\#}_{\Host}(d_2)},
\end{split}
\label{rho}
\end{equation}
where $d_{\min}=\min\{d_1,d_2\}$, $d_{\max}=\max\{d_1,d_2\}$.

The main assumption that we make in our experiments is the
following: we assume that the web host graph is obtained using a
\BO\ graph model, such as the graph $H$ described above. Under this
assumption, one can show that the cumulative characteristics of
the web host graph that we just defined have the following
properties.

\begin{itemize}
\item[1)] The function $\widetilde{\#}_{\Host}(d)$ is approximated well by
\begin{equation}
\label{f_estimate}
f_{a_1,b_1}(d)=b_1d^{-1-a_1}
\end{equation}
for some constants $a_1,b_1$ in an appropriate range of degrees. Note that the exponent in the power law is reduced by $1$ after the integration.

\item[2)] The function $\widetilde{\rho}_{\Host}(d_1,d_2)$ is
approximated well by\\
\begin{equation}
\label{rho_estimate}
g_{a_2,b_2}(d) = b_2(d_1+d_2)^{1-a_2}d_1^{a_2}d_2^{a_2}
\end{equation}
for some constants $a_2,b_2$ in an appropriate range of degrees. Note that the approximating function does not change after the integration (see Appendix).
\end{itemize}

Recall that under our assumption, we actually have $a_1 = a_2$. However, it is worth mentioning that just the fact that some graph satisfies both properties 1) and 2), does not automatically imply the equality $a_1 = a_2$. We explain it in detail in Section~\ref{Simulation}.

In our experiments, we estimate the constants $b_1,b_2,a_1,a_2$,
with the following results.

\begin{itemize}
\item[1)] The values of the approximating functions $f_{a_1,b_1}(d)$ and $g_{a_2,b_2}(d_1,d_2)$ are close enough to $\widetilde{\#}_{\Host}(d)$ and
$\widetilde{\rho}_{\Host}(d_1,d_2)$, respectively, for a sufficiently large range of degrees $d,d_1,d_2$.

\item[2)] The estimated values of $a_1$ and $a_2$ are very close
to each other, with relative difference only about $0.5\%$.
\end{itemize}
These facts make us believe that our main assumption about the realization of the \BO\ random graph with respect to the two quantitative aspects is reasonable.
The results are described in detail in Section~\ref{CumDistr}.

We describe and justify our method for estimation of the parameters $a_1$, $a_2$ in Section~\ref{Framework}. The experiments with simulated graphs confirm the validity of this method (Section~\ref{Simulation}).



\subsection{Data}
\label{Data}

All experiments are performed with the web host graph crawled in November 2011 by the major Russian search engine \texttt{yandex.ru}.
The robot is constantly crawling the web, collecting and updating web pages and links between them. From this data, cleaned from spam and duplicates, a web host graph can be constructed in the following way. Vertices of this graph correspond to owners. \emph{An owner} roughly corresponds to all pages downloaded by the robot at least once that belong to the same second level domain. (In some cases a second level domain is subdivided into several owners. Sometimes different second level domains are merged into a single owner.) An edge between two vertices-owners is drawn if there is a link from a page of one owner to a page of another owner. For the purposes of our work, we further simplify the graph, making it undirected and removing duplicate edges and self-loops. The web host graph constructed in this manner consists of 86.8~million vertices and 1.33~billion edges%
\footnote{To obtain the graph, please see \texttt{http://events.yandex.ru/events/publications/} or contact the authors.}%
.
We do not suspect any bias in the way this data is collected that may substantially affect our results.

\subsection{Framework for Parameter Estimation}
\label{Framework}

In Section~\ref{Preliminaries}, we already explained that the functions $f$ and $g$ defined by Equations~(\ref{f_estimate}) and (\ref{rho_estimate}) approximate $\widetilde{\#}_{\Host}$ and $\widetilde{\rho}_{\Host}$ for some appropriate values of the parameters $a_1,b_1$ and $a_2,b_2$, respectively. In this section we describe the method we use to optimize these parameters in order to obtain the best possible approximations.

Let $\Delta = \{[\alpha^k] : k \in \mathbb N\}$, where $\alpha = 1.01$, and $[\cdot]$ denotes the integer part of a number.

We use non-linear least squares method to minimize the overall deviation between empirical and theoretical functions over points from~$\Delta$.
Most authors fit the power law distribution to empirical data using a plain linear regression in log-log scale. Problems with this approach and reasons why it is not appropriate for fitting the degree distribution of a real graph have already been discussed extensively~\cite{PowerLaw}. In our case, we see the following additional reasons not to use this method.
\begin{itemize}

\begin{figure}[!ht]
\centering
\includegraphics[width=0.4\textwidth]{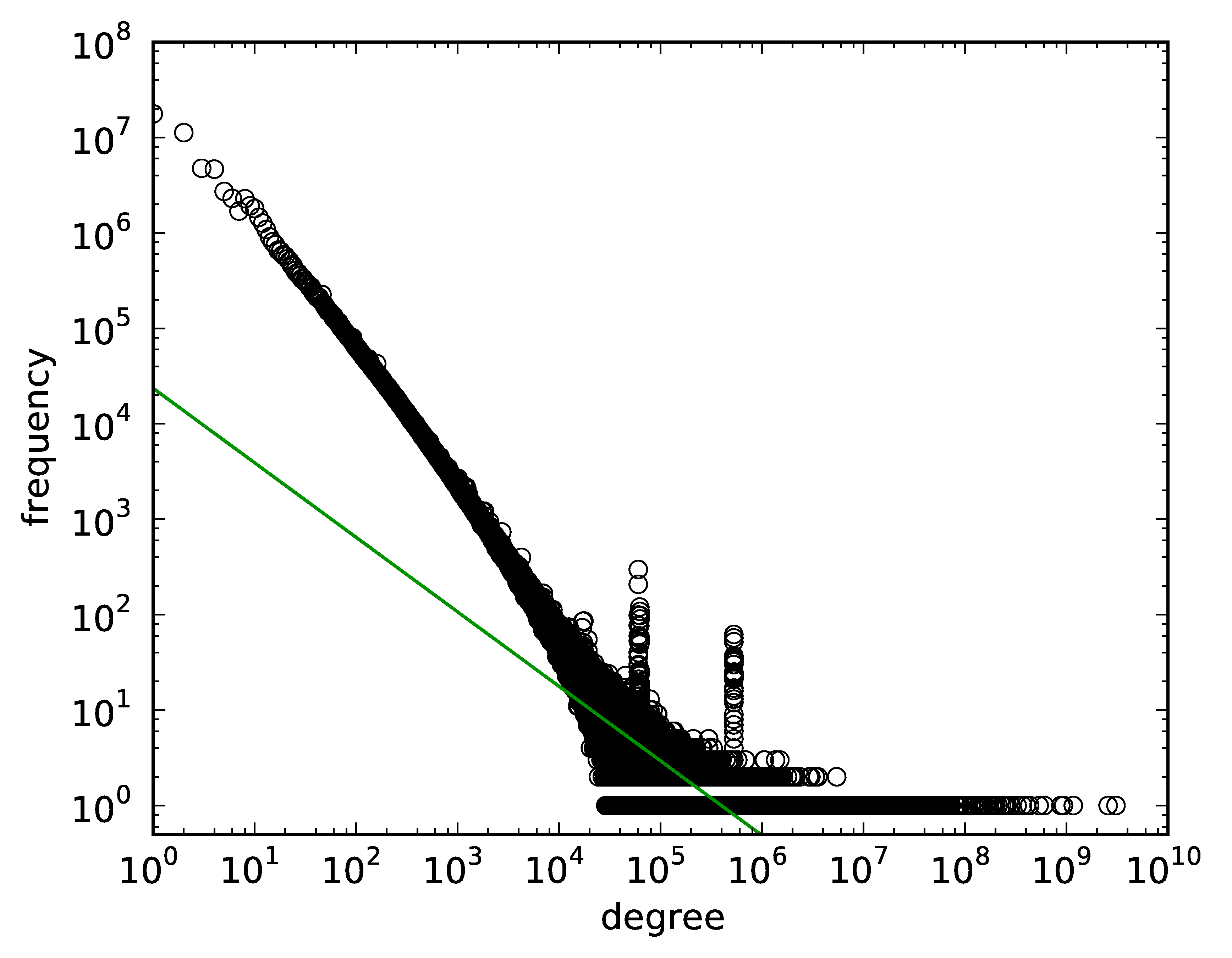}
\caption{Degree distribution of the web host graph (for each degree, the number of vertices having this degree is represented with a black circle) and approximation using linear regression (green) in logarithmic scale.}
\label{fig:host}
\end{figure}

\begin{figure}[!ht]
\centering
\begin{tabular}{ll}
\includegraphics[width=0.4\textwidth]{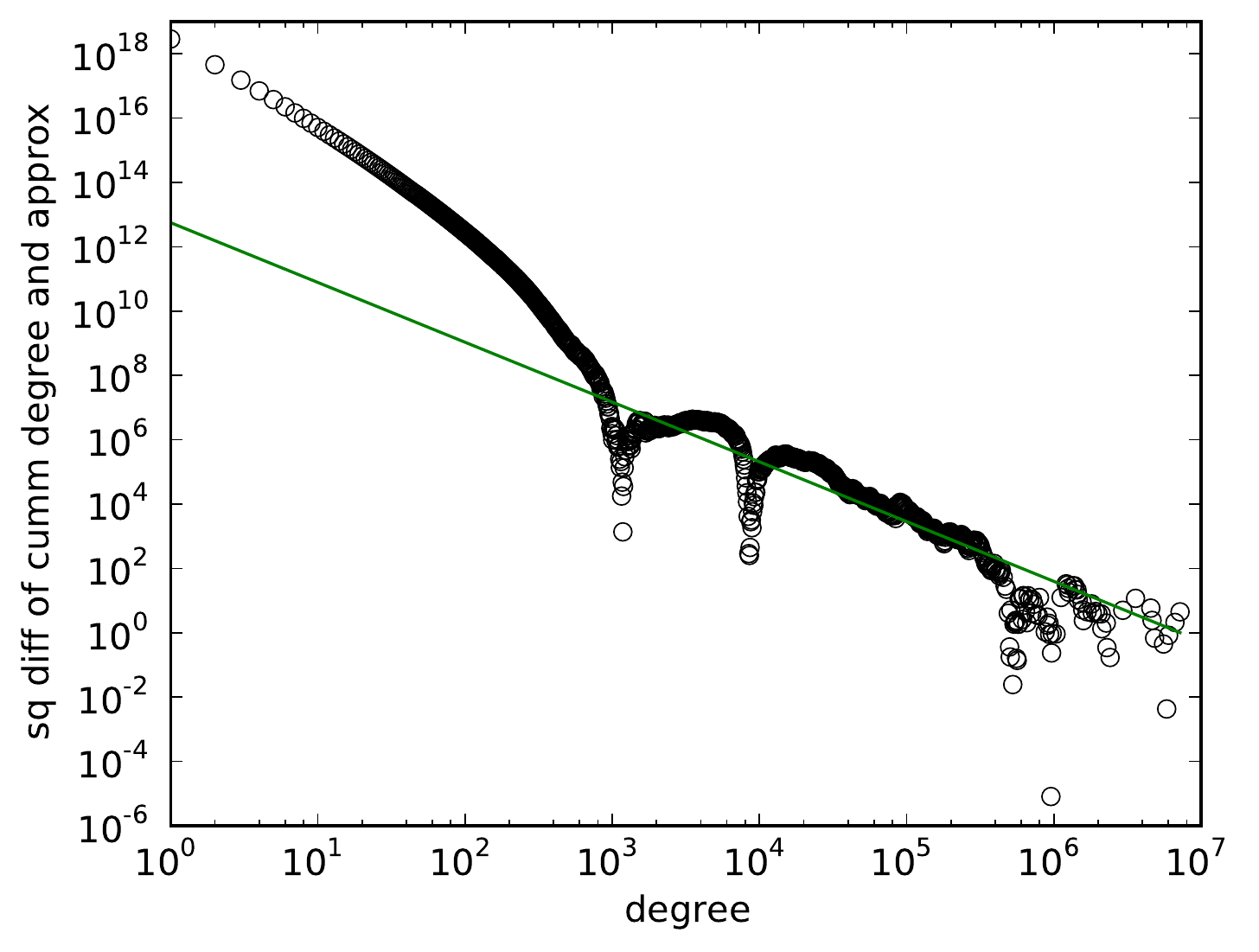}
&
\includegraphics[width=0.4\textwidth]{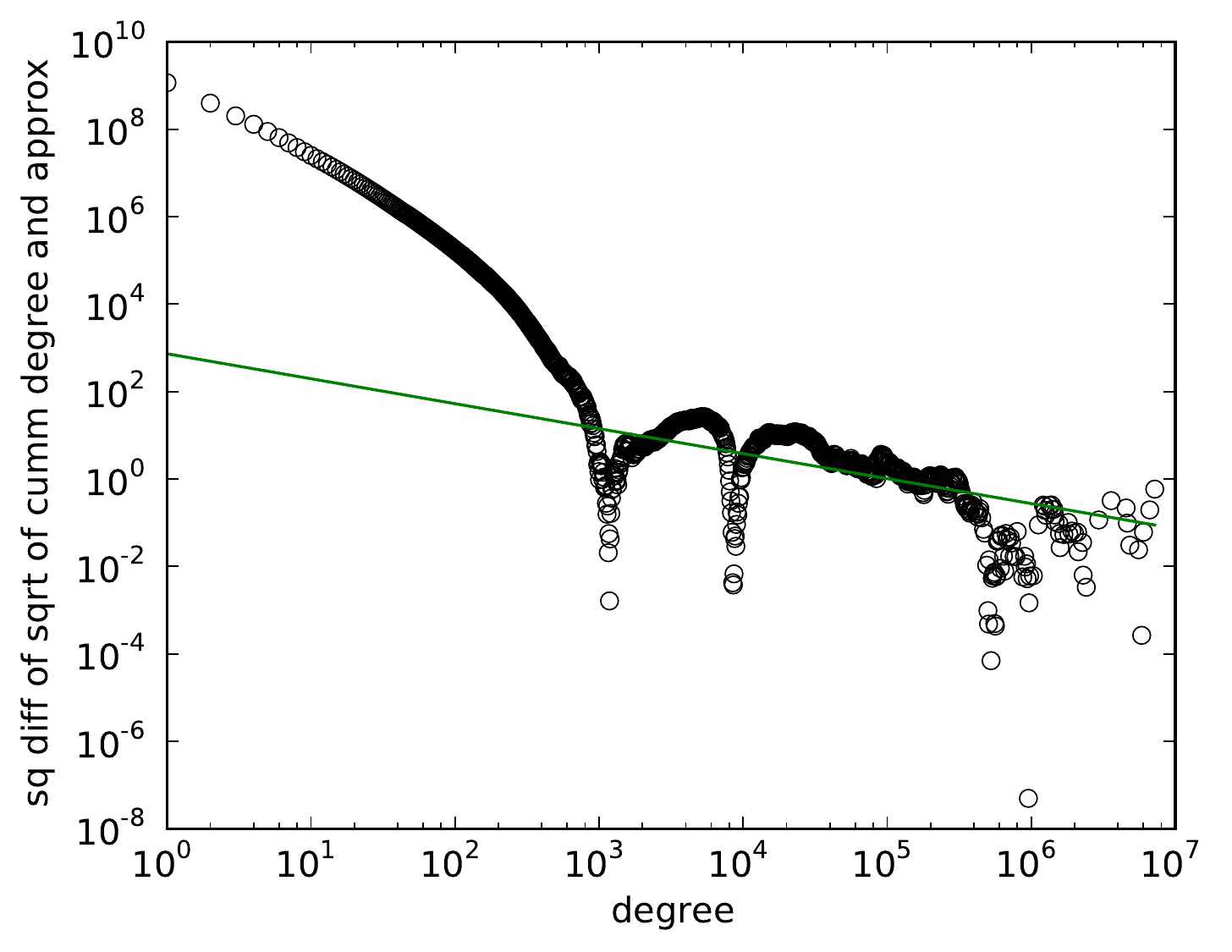}
\\
(a) & (b)
\end{tabular}
\caption{(a) Squared deviation between cumulative degree distribution
and approximation using our method. (b) Squared deviation between \emph{square roots} of cumulative degree distribution and approximation using our method. In both cases, linear regression for the range $[10^{2.9},10^{5.9}]$ is shown for convenience.}
\label{fig:dispersions}
\end{figure}

\item[1)] Empirical argument: Fig.~\ref{fig:host} illustrates that the linear regression for $\log(\#_{\Host}(d))$ is a pretty bad approximation.

\item [2)] It can be shown that (assuming the \BO\ model) the variance and the mean of $\widetilde{\#}_{\Host}(d)$ have the same order of growth as $d$ grows, and therefore the variance of $\sqrt{\widetilde{\#}_\Host(d)}$ can be bounded by a constant. This means that the following objective function
\begin{equation}
\label{eq:objective-degrees}
\frac{1}{|D_1|}
\sum_{d\in D_1}\left(\sqrt{\widetilde{\#}_{\Host}(d)}-\sqrt{f_{a_1,b_1}(d)}\right)^2,
\end{equation}
where $D_1 \subset \Delta$ is a certain range of degrees,
is much more appropriate for optimization, as in this case contributions of different summands are better calibrated. To illustrate the validity of this argument empirically, we plot the distribution of $\widetilde{\#}_\Host(d) - f_{a_1,b_1}(d)$ (Fig.~\ref{fig:dispersions}a) and $\sqrt{\widetilde{\#}_{\Host}(d)}-\sqrt{f_{a_1,b_1}(d)}$ (Fig.~\ref{fig:dispersions}b) for parameters of $a_1,b_1$ estimated in Section~\ref{CumDistr}.

\item[3)] The linear regression is not applicable to estimation of parameters $a_2,b_2$, for function $g$ can not be represented in a linear form.
Moreover, we are again able to show that the variance and the mean of the empirical probability of an edge are of the same order.
\end{itemize}

In accordance with 3), we estimate parameters $a_2,b_2$ minimizing
the following objective function
\begin{equation}
\label{eq:objective-edges}
\frac{1}{|D_2|}\sum_{(i,j) \in D_2}\left(\sqrt{\widetilde{\rho}_{\Host}(i,j)}-\sqrt{g_{a_2,b_2}(i,j)}\right)^2,
\end{equation}
where $D_2 = \{(d_1,d_2)\in D_1^2 : d_1/d_2 > 10\}$, and $D_1$ is the degree range chosen for estimating $a_1,b_1$ (to be determined later). Note that we introduce the restriction on $d_1/d_2$ in accordance with Theorems \ref{BO-edges-mean} and~\ref{BO-edges-var} that give a good estimation for $\widetilde{\rho}$ only for sufficiently large $d_1/d_2$ (see Section~\ref{EdgesTheory}). The value $C = 10$ was chosen manually.

We minimize the objective functions (\ref{eq:objective-degrees}) and (\ref{eq:objective-edges}) using the Gauss--Newton algorithm for a non-linear least squares optimization problem (e.g., see~\cite{Econometrics}). Varying the degree range $D_1$, we examined the product of the resulting optimized objectives (\ref{eq:objective-degrees}), (\ref{eq:objective-edges}) and chose $D_1 = [10^{2.9},10^{5,9}]$ as the range of length 3 (in the logarithmic scale) with the minimal value of the product.
The choice of the degree range for our approximations is further justified empirically in our observations on the deviations (Fig.~\ref{fig:dispersions}). For ranges of larger lengths, the optimized product of objectives starts to grow substantially.

In the next subsection we describe the results of our experiments.

\subsection{Estimation for Empirical Cumulative Degree and Edge Distributions}
\label{CumDistr}

In this section we discuss the results of the two estimation methods  for the parameter $a$ described in Section~\ref{Framework}.

Table~\ref{table:edges} shows the estimate of $a_2$ we obtained deriving the best fit of $g_{a_2,b_2}$ to the empiric conditional probability $\widetilde{\rho}_{\Host}(d_1,d_2)$ that a pair of vertices $v_1,v_2$ forms an edge in the web host graph given that $\max(\deg(v_1), \deg(v_2)) > \max(d_1,d_2)$ and $\min(\deg(v_1), \deg(v_2)) > \min(d_1,d_2)$ (see Equation~(\ref{rho}) for the definition).

\begin{table}[!ht]
\centering
\begin{tabular}{|c||c|c|}
  \hline
  parameter & degree distribution & edge distribution \\
  \hline
  \hline
  $a$ &  $a_1 = 0.2762$ & $a_2 = 0.2774$ \\
  \hline
  $\sigma$ & 2.631 & 0.0599 \\
  \hline
  $\sigma_s$ & 0.005666 & $8.518 \cdot 10^{-6}$ \\
  \hline
\end{tabular}

%

\caption{\small Results of the approximation of the cumulative distribution for degrees from the interval $D_1 = [10^{2.9},10^{5.9}]$ and for edges between vertices with degrees from the domain $D_2$ for the web host graph (see Sections~\ref{Preliminaries} and \ref{Framework} for details).}
\label{table:edges}
\end{table}

\begin{figure}[!ht]
\centering
\includegraphics[width=0.4\textwidth]{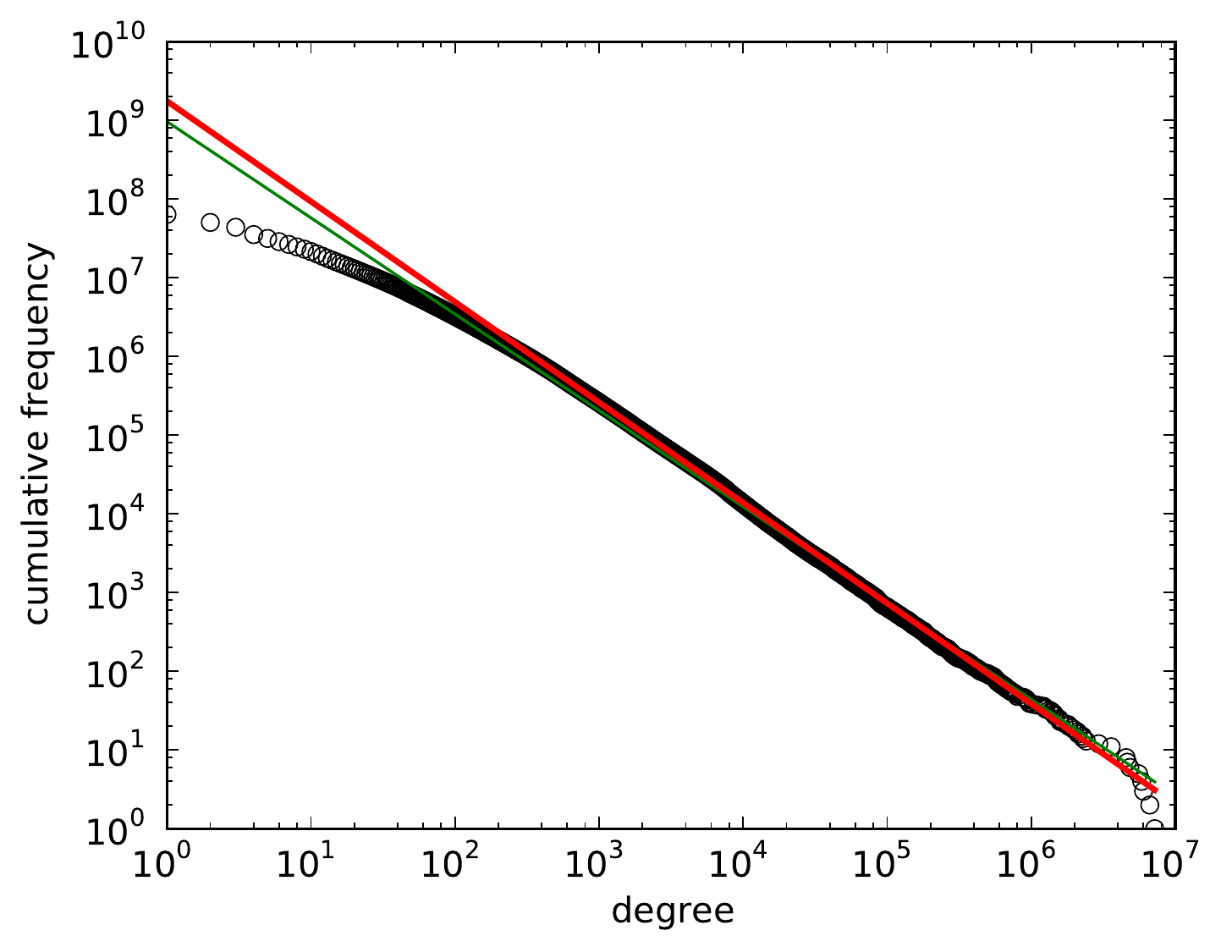}
\caption{Cumulative degree distribution (black) and
approximation using our method (red) in logarithmic scale. For
comparison, the result of approximation using linear regression in
log-log coordinates is shown (green).}
\label{fig:degrees}
\end{figure}

We measure the accuracy of the estimation of $a_2$ employing
bootstrapping in the following way. We sample the set of edges of
the same size as originally, choosing each edge uniformly at
random from the collection of all edges, with replacement. For
each sample, we substitute the empirical function
$X_{\Host}(d_1,d_2)$ with that for the sampled set, refresh
$\widetilde{\rho}_{\Host}(d_1,d_2)$ according to (\ref{rho}) and
apply the estimating method described in Section \ref{Framework}. Applying the described procedure 1000 times independently, we obtain one estimate for $a_2$ for each edge
sampling. The normalized sum of squared deviations between these
1000 estimates and the one obtained from the
initial dataset is shown in Table~\ref{table:edges} as
$\sigma_s^2$. We denote the normalized sum of squared deviations
between $g_{a_2,b_2}(d_1,d_2)$ and
$\widetilde{\rho}_{\Host}(d_1,d_2)$ in the domain $D_2$ by $\sigma^2$.

\begin{figure*}[!ht]
\centering
\includegraphics[width=0.4\textwidth]{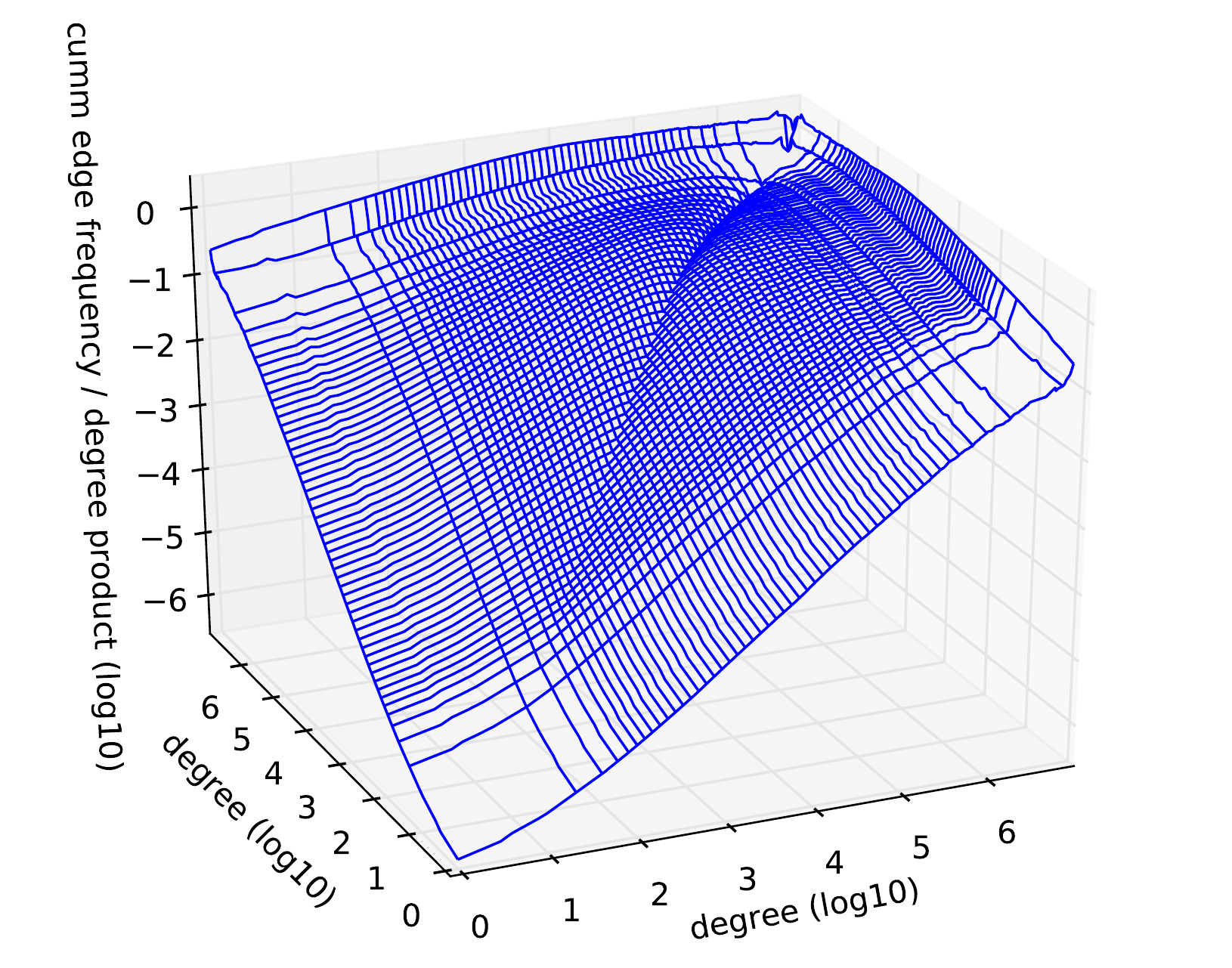}
\includegraphics[width=0.4\textwidth]{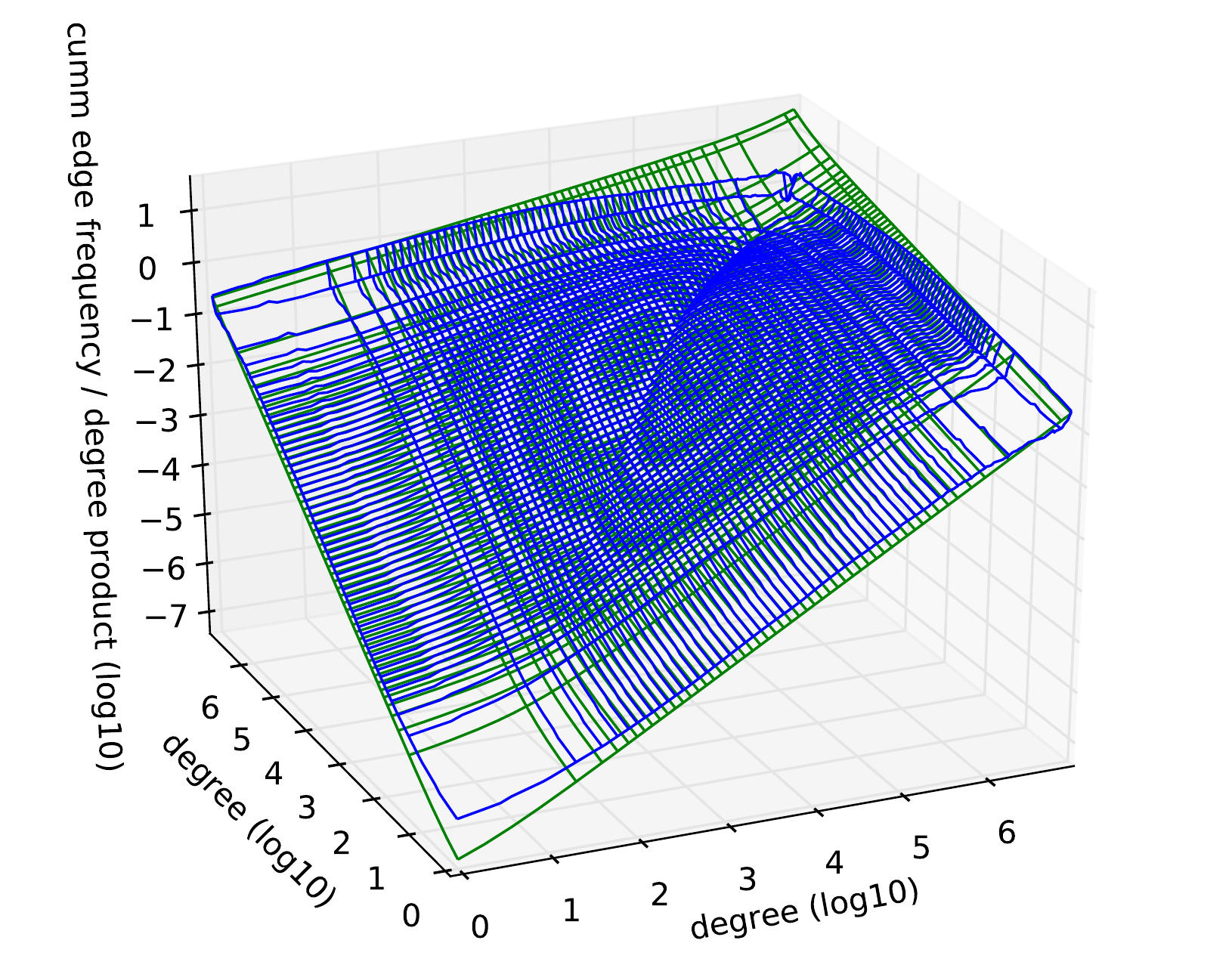}
\includegraphics[width=0.4\textwidth]{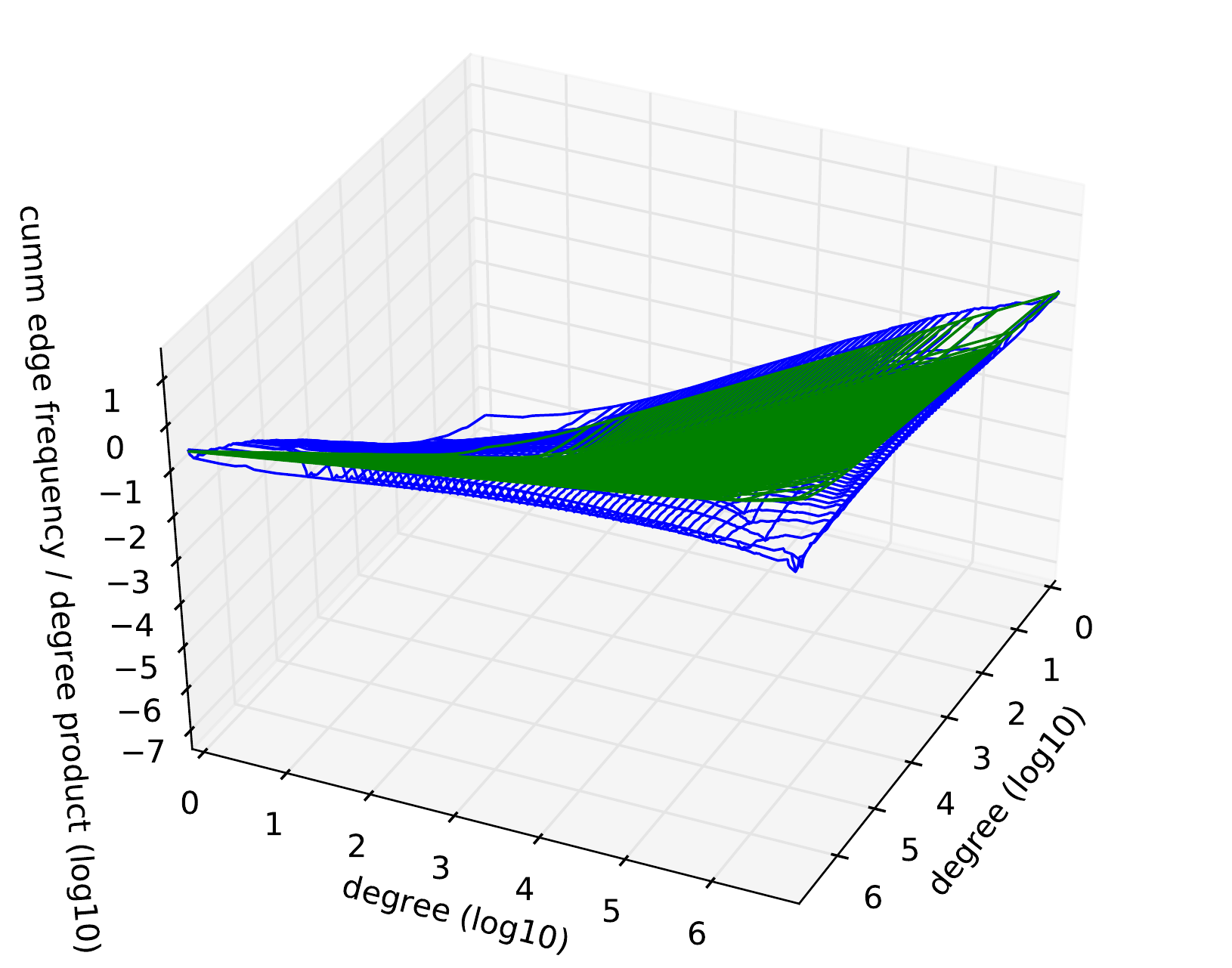}
\caption{Cumulative edge distribution (blue) and
approximation using our method (green) in logarithmic scale (axes
are labeled with $\log_{10}$ of the values, pictures differ only in the view angle).}
\label{fig:edges}
\end{figure*}

For the chosen range of degrees $D_1 = [10^{2.9}, 10^{5.9}]$, we obtain $a_2 = 0.2774$ and $b_2 = 8.331\cdot 10^{-4}$.
The results of this approximation are shown on
Fig.~\ref{fig:edges}. We observe a very good fit of approximation
with the data. Note that due to the term $(d_1 + d_2)^{1 - a}$
predicted theoretically, the approximation was even able to
capture a concave area around the diagonal $d_1 = d_2$. This would
not be possible with a simpler approximation of the form $d_1^a
d_2^a$.

The result of estimation of $a_1$ that we obtained approximating
$\widetilde{\#}_{\Host}$ by the function $f_{a_1,b_1}$ in the
range of degrees $D_1$ is also shown in Table~\ref{table:edges}. We
also measure the estimation accuracy using bootstrapping, sampling
with replacement 1000 sets of vertices, applying our method to the
corresponding degree distribution and obtaining 1000 values of
estimates. 
The normalized sum of squared deviations between
$f_{a_1,b_1}(d_1,d_2)$ and $\widetilde{\#}_{\Host}(d_1,d_2)$ is
denoted by $\sigma^2$ as well.

We want to stress that surprisingly we obtained the same value $a\approx 0.27$ of the parameter approximating independently the degree and the edge distributions. This is a double evidence that the \BO\ model is good for the web host graph. In the next section, we further support this claim, comparing it with other models.



\section{Experiments on Simulated Graphs}
\label{Simulation}

Here we describe the results of our experiments with graphs artificially generated in various random graph models. We have two goals: to demonstrate that for a random graph with the power law degree distribution the probability of an edge between vertices of given degrees is not determined by the exponent in the power law, and to show that the \BO\ model has the best approximation to the web host graph as compared with other models.

\begin{figure}[!ht]
\centering
\begin{tabular}{ll}
\includegraphics[width=0.4\textwidth]{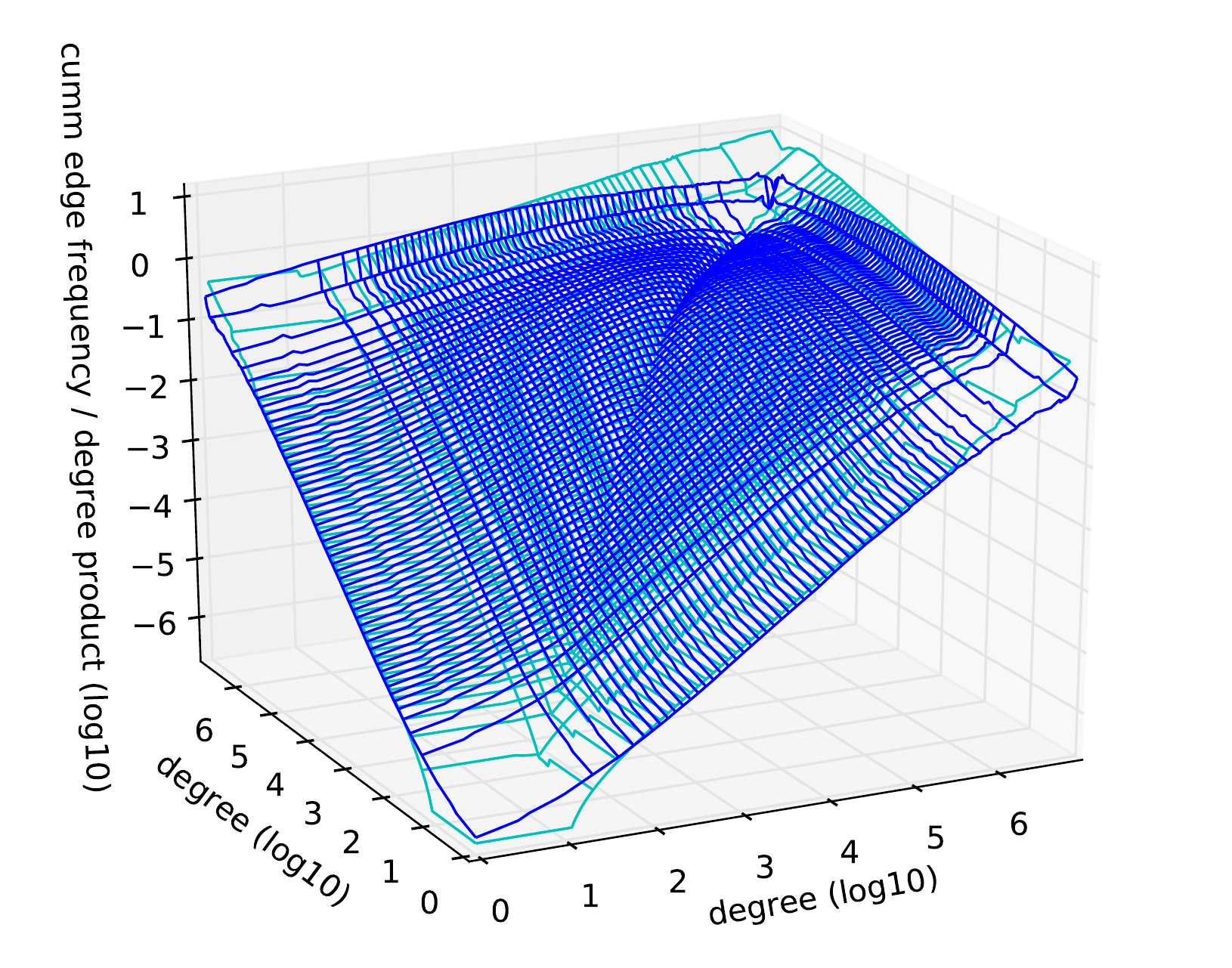} &
\includegraphics[width=0.4\textwidth]{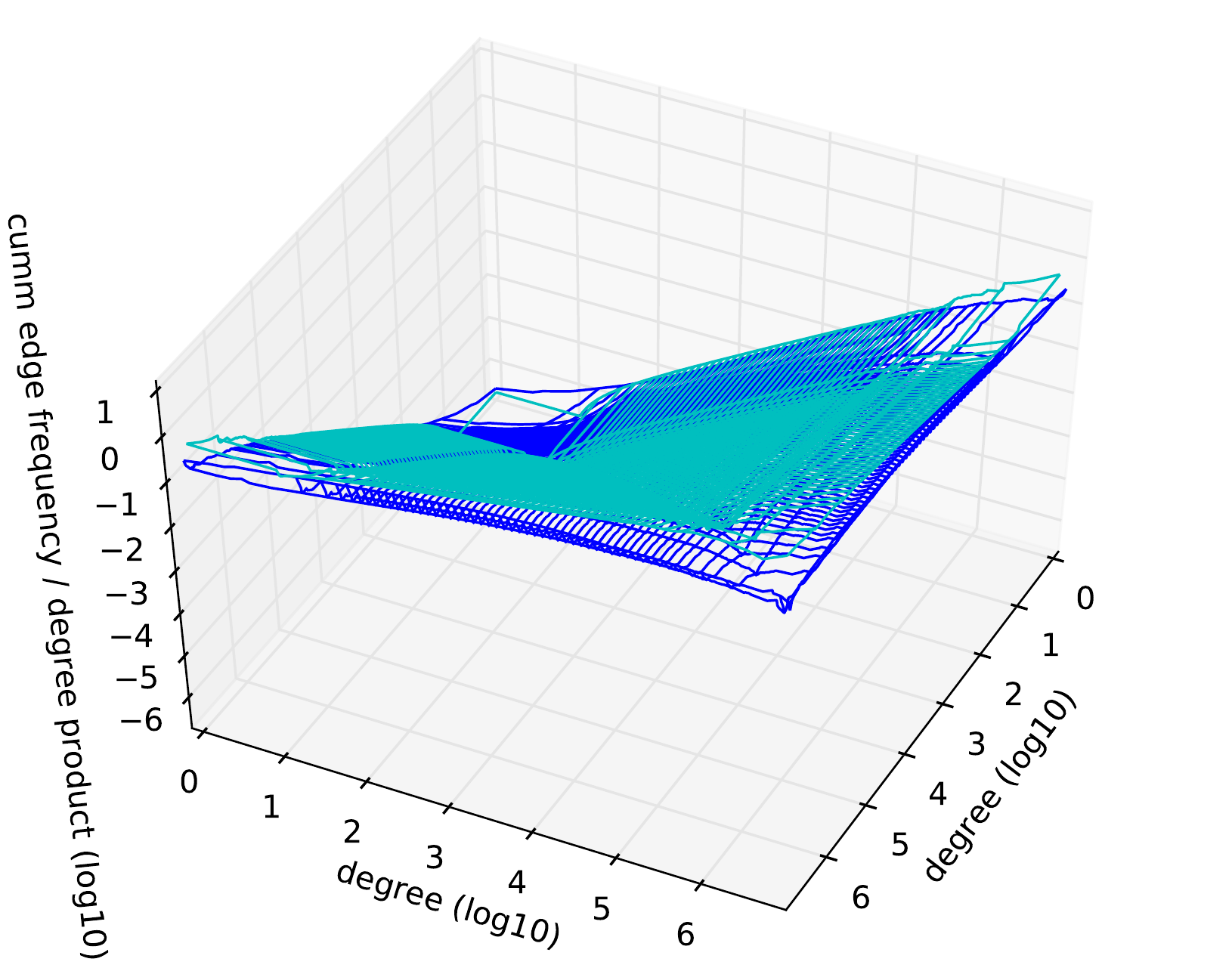} \\
\end{tabular}
\caption{Cumulative edge distributions for the
web host graph (blue) and for the \BO\ simulated graph (cyan) in logarithmic scale (axes are labeled with $\log_{10}$ of the values, pictures differ only in the view angle).}
\label{fig:simulatedBO}
\end{figure}

\begin{figure}[!ht]
\centering
\includegraphics[width=0.4\textwidth]{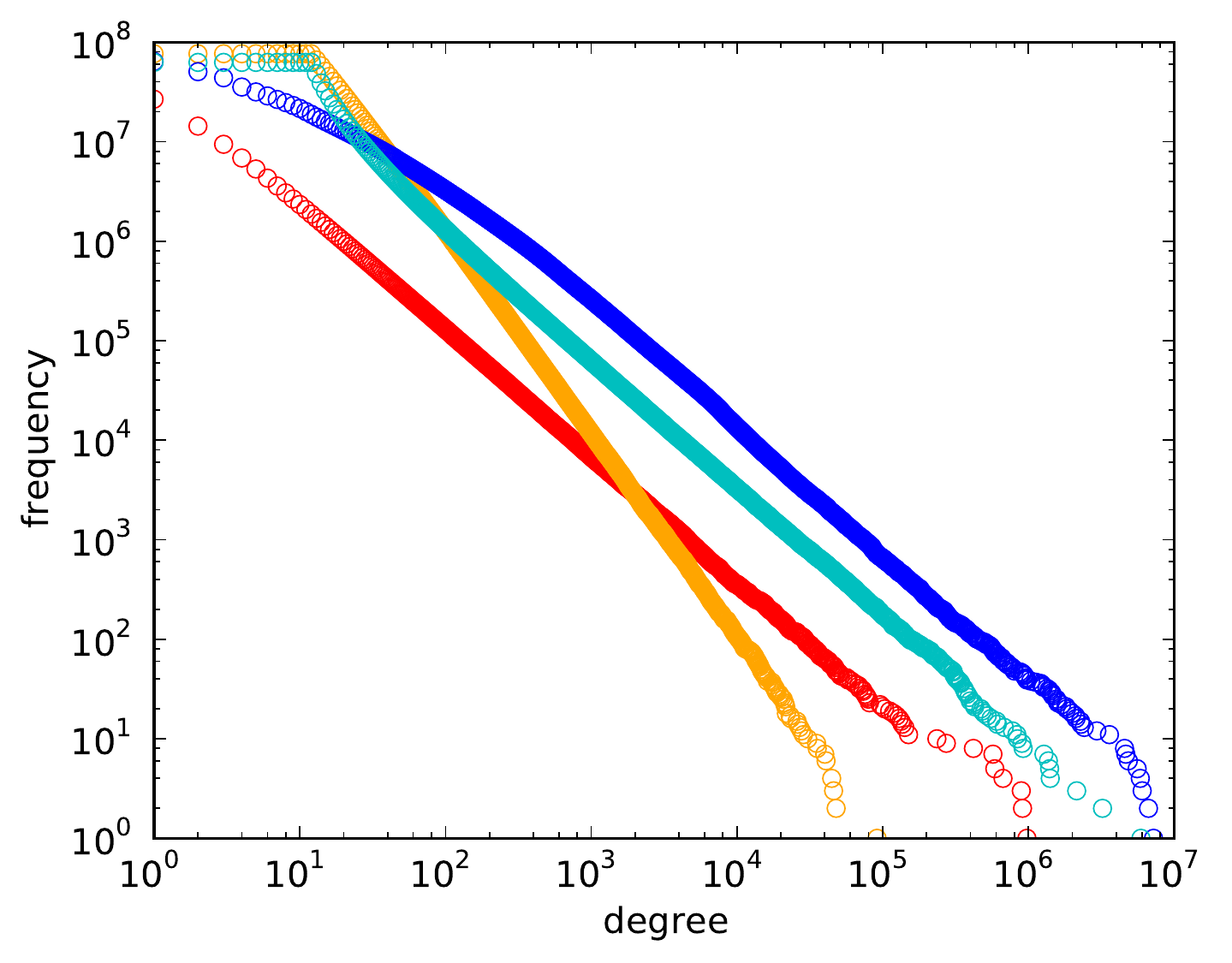}
\caption{Cumulative degree distributions for the web host graph (blue), the BO simulated graph (cyan), the GDS simulated graph (red), and the HK simulated graph (orange) in logarithmic scale.}
\label{fig:simulatedOthersDegrees}
\end{figure}

First of all, we generate ten samples of the \BO\ (BO) random graphs with 86.8M vertices with $a = 0.276$ and $m = 12$ (close to the ratio of the number of edges and the number of vertices observed in the actual web host graph). The cumulative degree and edge distributions of one resulting graph are shown on Fig.~\ref{fig:simulatedOthersDegrees} and~\ref{fig:simulatedBO}, respectively, in comparison with those for the web host graph. In both cases, we observe a strong fit, recapitulating the results from Section~\ref{CumDistr} (compare with Fig.~\ref{fig:edges}).

\begin{figure}[!ht]
\centering
\includegraphics[width=0.4\textwidth]{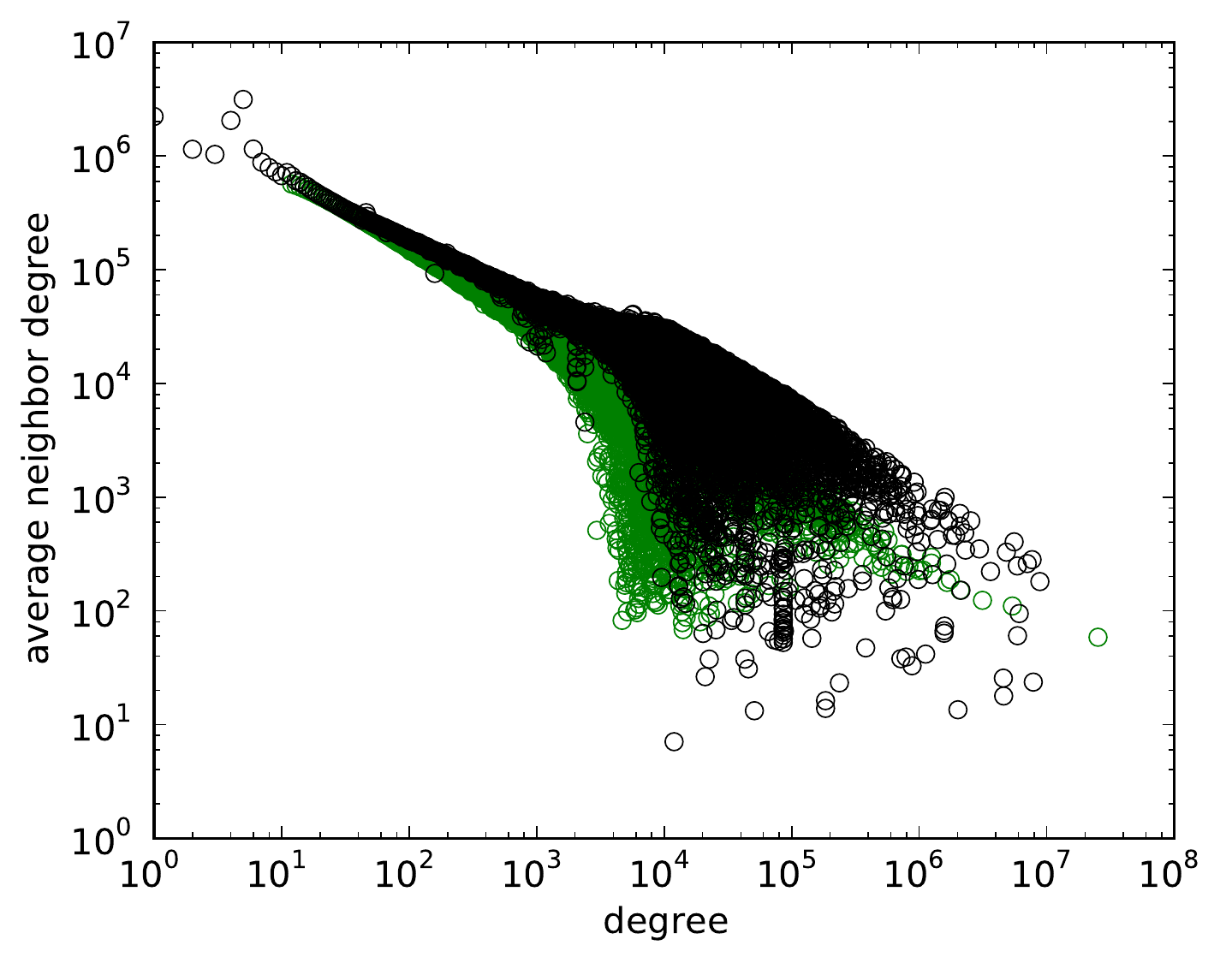}
\caption{Average degree of a neighbor of a vertex depending on the degree of this vertex for the real web host graph (black) and a sample generated by the \BO\ model (green) with $a = 0.276$ (corresponding to the best approximation).}
\label{fig:assortativity}
\end{figure}

Fig.~\ref{fig:assortativity} compares the function $d_{nn}$, average degree of a neighbor, for the web host graph and a sample generated in the \BO\ model with $a = 0.276$ that corresponds to the best approximation by the model. As expected, the two distributions are very close to each other. Interestingly, even fluctuations of the two are very similar.

In addition to the \BO\ model, we consider two other random graph models: the configuration model (GDS) and the Holme--Kim model (HK).

The first model chooses from all graphs with a specified fixed degree sequence uniformly at random~\cite{config}. For our experiment, we generate a sequence of 86.8M numbers following the power law distribution with the exponent $-2.276$ and use this distribution as a degree sequence in the model.
Then we generate five samples of random graphs in this model
with 86.8M vertices and 128M edges using a simple simulation algorithm~\cite{config}. The degree distribution of the resulting graph follows the power law by construction.

The second model is based on the idea of preferential attachment with triad formation steps in the graph construction process~\cite{HK}. We generate nine samples of random graphs with 86.8M vertices and 1B edges. Degree distribution of the resulting graph follows the power law with the exponent $-3$.

\begin{figure}[!ht]
\centering
\begin{tabular}{ll}
\includegraphics[width=0.4\textwidth]{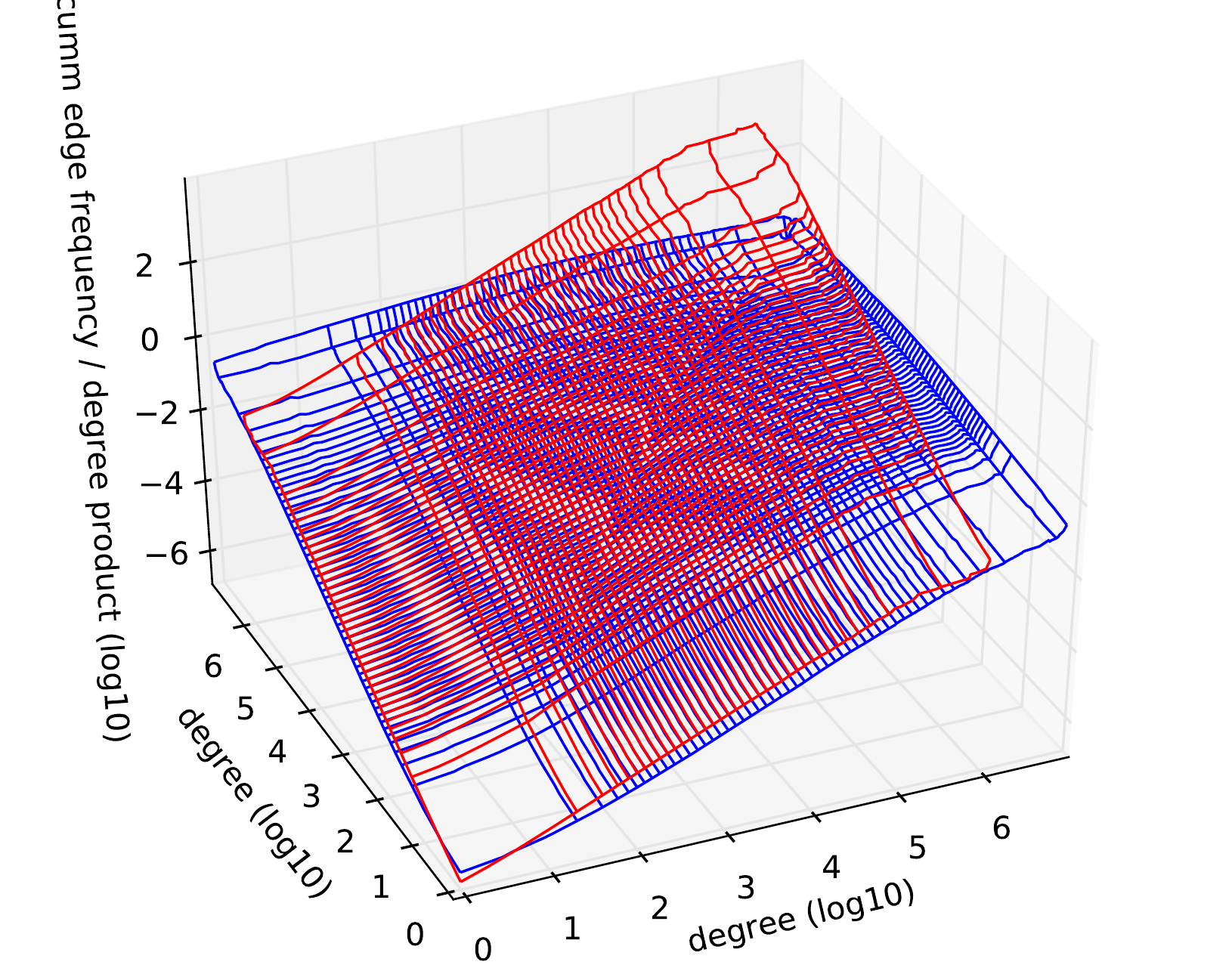} &
\includegraphics[width=0.4\textwidth]{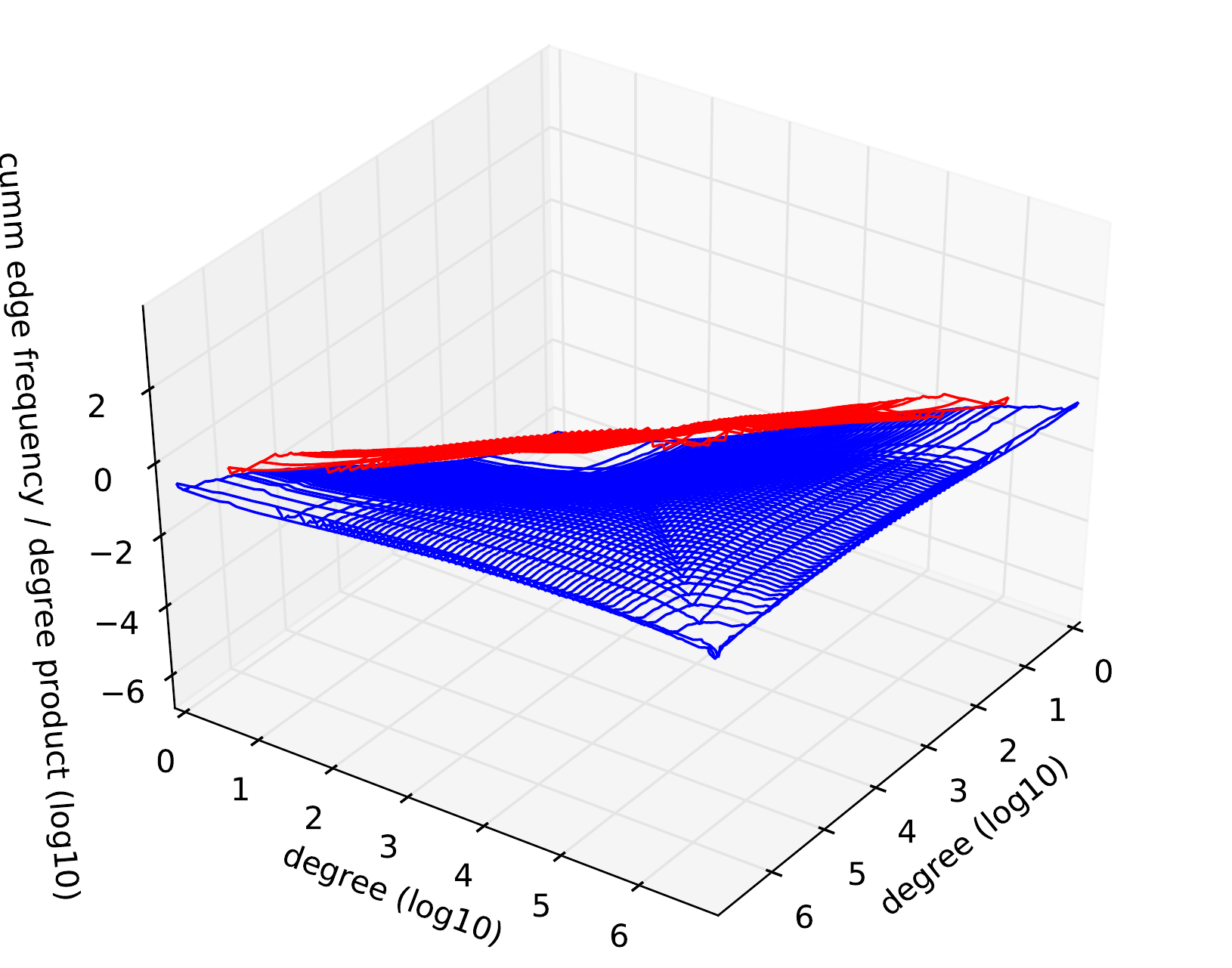} \\
\includegraphics[width=0.4\textwidth]{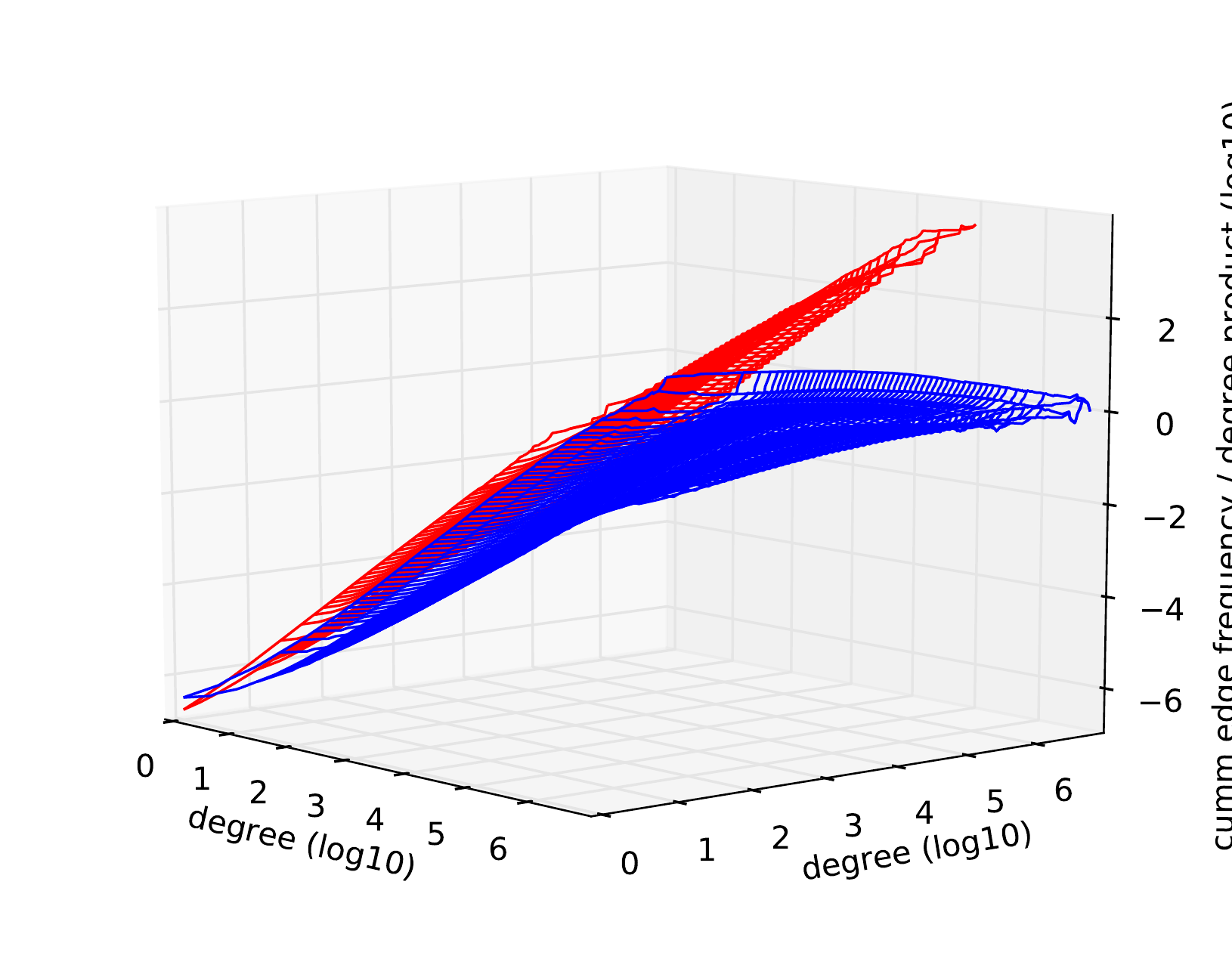} &
\includegraphics[width=0.4\textwidth]{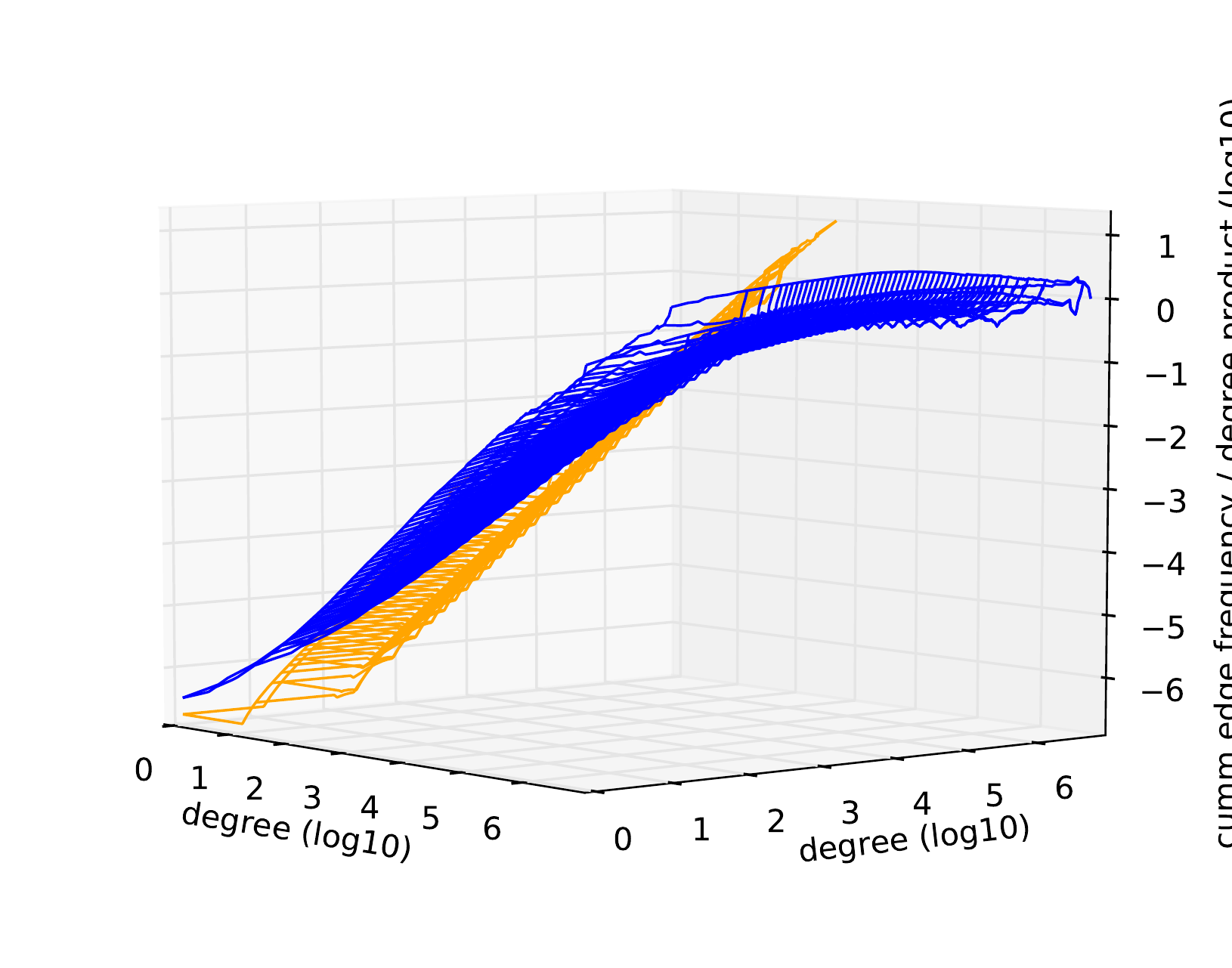}
\end{tabular}
\caption{Cumulative edge distributions for the
web host graph (blue), the GDS simulated graph (red), and the HK simulated graph (orange) in logarithmic scale. Pictures for the GDS model differ only in the view angle.}
\label{fig:simulatedOthers}
\end{figure}

The degree and the edge distributions for a single sample from both models in comparison with those for the web host graph are shown on Fig.~\ref{fig:simulatedOthersDegrees} and~\ref{fig:simulatedOthers}, respectively.

For each of the simulated graphs, we apply exactly the same two approximation procedures as described in Section~\ref{Framework} and previously applied to the web host graph. Table~\ref{table:generated} shows the results: $v$ and $e$ are the number of vertices and edges in the sample graphs,
$a_1$ and $a_2$ are the parameters of the best fit for degree and edge distributions, respectively. Note that the algorithm diverges for edge distribution approximation of the HK model, and the value of $a_2$ is not defined in this case. We also show the standard deviation of the obtained estimates of $a_1$ and $a_2$ over the several samples of the model. The GDS model has a fixed degree distribution that results in always the same estimate of~$a_1$.

Not surprisingly, the approximation algorithm extracts the parameters $a_1$ and $a_2$ planted in the sample of the BO model with high accuracy, as it is the underlying assumption of this algorithm that the graph is modeled by the \BO\ model.

\begin{table*}[!ht]
\centering
\begin{tabular}{|r||c|c|}
  \hline
  model & parameters & estimates \\
  \hline
  \hline
  BO &  $v = 8.68 \cdot 10^7$, $e=1.04\cdot 10^9$
  &
  $a_1=0.289\pm 0.0033$, $a_2=0.274\pm 0.0038$ \\
  \hline
  GDS & $v = 8.68 \cdot 10^7$, $e=1.26\cdot10^8$
  & $a_1 = 0.29 \pm 0$, \hspace{6.5mm} $a_2=1.053\pm 0.00048$ \\
  \hline
  HK & $v = 8.68 \cdot 10^7$, $e = 1.04\cdot 10^9$ & $a_1 = 1.06\pm 0.0088$,  \hspace{16mm} $a_2 = $ n/a \\
  \hline
\end{tabular}
\caption{Results of the approximation of the cumulative distributions
of degrees from the interval $D_1=[10^{2.9},10^{5.9}]$ and edges between vertices with degrees from the interval $D_1$ for generated graphs (see Sections~\ref{Preliminaries} and \ref{Framework} for details). Number of vertices and edges in graphs are shown as $v$ and $e$, respectively. Results of the approximation using the method described in Section~\ref{Framework}, are shown as $a_1$ and $a_2$.}
\label{table:generated}
\end{table*}

Although all generated graphs have the power law degree distribution, only the \BO\ graph has the edge distribution close to that observed in the real web host graph.

\section{Conclusion}
\label{Conclusion}
In this paper we study the degree and edge distributions of the web host graph. We compare it with the Buckley--Osthus model of random graphs and find that the model agrees with the real data. More precisely, we use two different approaches to estimate the initial attractiveness parameter $a$ assuming the web host graph is generated in the \BO\ model. In two different independent attempts, we compare the distribution of the number of edges between vertices with respect to their degrees and the degree distribution in the real graph with theoretical predictions for the \BO\ model. The values of $a$ obtained with two methods are very close to each other, and therefore we conclude that the web host graph is very similar to the \BO\ random graph with this particular value of~$a$.

Besides our results being interesting on their own, we believe they may potentially be related with real world problems of practical interest.

One example of such a relation may be the work of Y.~Lu et al.~\cite{Lu} that made use of the power law degree distribution in the webgraph and proposed the algorithm PowerRank, an improvement over PageRank. We may expect that further empirical and theoretical studies of graphs representing the Internet may help progress in other tasks related with search and in particular with ranking and crawling.

It has been argued that the web contains many communities, sets of pages or hosts that are in particular characterized by abnormally high density of links between them~\cite{Bonato2005,Gibson,Kleinberg}. In this respect, understanding how edges are distributed in the graph may potentially be useful for algorithms detecting and testing such communities, providing a better description of expected background that prospective communities may be compared against. We expect that theoretical and empirical results in the direction presented in this paper may prove useful for these problems.

One can imagine a lot of directions for future work related with our results, both theoretical and practical.

It would be interesting to continue to study the \BO\ random graph model, as well as other models, and extend theoretical knowledge of their properties. For the first time we described the distribution of edges between vertices given their degrees in a real Internet graph. Now it is interesting to compare different models with respect to this property, and our techniques may be useful.

Even though we showed a good correlation of the model with real data, we had to simplify the data in certain important aspects. It would be interesting to generalize existing random graph models or probably to develop new ones that could model graphs closer to the reality: with multiple edges, directed, hierarchical, dynamically evolving with time. In particular, the clustering coefficient of a \BO\ graph still significantly differs from the one in the reality. However, some of the aspects of the \BO\ model may be promising.

It would definitely be interesting to develop and test the aforementioned and similar ideas of applications to ranking, crawling, and community detection. We strongly believe that deeper and broader theoretical results on models of Internet graphs coupled with empirical observations of certain characteristics of real such graphs may lead to practical applications and insights.

\appendix

\section{Proof of Proposition~1} 
\label{ProofProposition}

We can estimate the expectation of the number of loops in the following way:
$$
E N(\text{loops in }H_{a,m}^n) = O\left( \sum_{i=1}^n \frac{1}{i}
\right) =  O\left(\ln n\right).
$$
To estimate the number of multiple edges we should take into account that we have no vertices of degrees greater than $2mn$ in $H_{a,m}^n$.
Also (using the same ideas as in the proof of Theorem 3) it can be shown that $E\#_a(d,i) = O \left(\frac{i}{d^{2+a}}\right)$. Therefore
$$
E N(\text{multiple edges in }H_{a,m}^n)  =
$$
$$
= O\left(\sum_{i=1}^n \sum_{d=1}^{2mi} E\#_a(d,n)\left( \frac{d-1+a}{(a+1)i} \right)^2 \right) =
$$
$$
= O\left(\sum_{i=1}^n \sum_{d=1}^{2mi} \frac{i}{d^{2+a}} \frac{d^2}{i^2} \right) =
O\left(n^{1-a}\right).
$$

\section{Proof of the theoretical approximation in Equation~(7)
}
\label{ProofEstimate}

Here we prove the theoretical approximation from Equation~(\ref{rho_estimate}) for the empiric conditional probability $\widetilde{\rho}_{\Host}(d_1,d_2)$.

First, for sufficiently large $d_1/d_2$, we obtain the following approximate formula using the estimations (\ref{v_estimate}) and (\ref{X_estimate}):

\begin{equation}
\label{rho_HOST_estimate}
\widetilde{\rho}_{\Host}(d_1,d_2)\approx \frac{b_2\sum_{i\geq j,\,i>d_1, j>d_2} (i + j)^{1-a_2}(ij)^{-2}}{b_1^2\sum_{i>d_1} i^{-2-a_2}\sum_{j>d_2} j^{-2-a_2}}.
\end{equation}

For $d_1/d_2$ large enough, the numerator of the right-hand side of (\ref{rho_HOST_estimate}) equals

$$
 \sum_{i>d_1\geq j >d_2} b_2(i +
 j)^{1-a_2}(ij)^{-2}+
 \sum_{i\geq j>d_1} b_2(i +
 j)^{1-a_2}(ij)^{-2}\approx
$$
$$
 c_1(d_1+d_2)^{1-a}(d_1d_2)^{-1}+c_2(d_1)^{-1-a}\approx c_1(d_1+d_2)^{1-a}(d_1d_2)^{-1}
$$
for some constants $c_1,c_2$. Estimating the denominator of the right-hand side of (\ref{rho_HOST_estimate}) by $c(d_1d_2)^{-1-a_2}$, we get $\widetilde{\rho}_{\Host}(d_1,d_2)\approx g_{a_2,b_2}(d_1,d_2)$, where $g_{a_2,b_2}$ is defined by (\ref{rho_estimate}).

\end{document}